\documentclass[a4paper,11pt]{article}
\pdfoutput=1 

\usepackage{jinstpub} 

\title{Design and Manufacture of the RF Power Supply and RF Transmission Line for SANAEM  Project Prometheus}

\author[a,b,1]{G. Turemen,\note{Corresponding author.}}
\author[c]{S. Ogur}
\author[d]{F. Ahiska}
\author[a,b]{B. Yasatekin}
\author[e]{E. Cicek}
\author[f]{A. Ozbey}
\author[b]{I. Kilic}
\author[g]{G. Unel}
\author[b]{A. Alacakir}

\affiliation[a]{Ankara University, Department of Physics, Ankara, TURKEY}
\affiliation[b]{TAEK, Saraykoy Nuclear Research and Training Center, Ankara, TURKEY}
\affiliation[c]{Bogazici University, Department of Physics, Istanbul, TURKEY}
\affiliation[d]{EPROM Electronic Project \& Microwave Ind. and Trade Ltd. Co., Ankara, TURKEY}
\affiliation[e]{Gazi University, Department of Physics, Ankara, TURKEY}
\affiliation[f]{Istanbul University, Aircraft Technology Program, Istanbul, TURKEY}
\affiliation[g]{University of California at Irvine, Department of Physics and Astronomy, Irvine, USA}

\emailAdd{gorkem.turemen@ankara.edu.tr}

\abstract{A 1-5~MeV proton beamline is being built by the Turkish Atomic Energy
Authority in collaboration with a number of graduate students from different universities. The primary
goal of the project, is to acquire the design ability and manufacturing capability
of all the components locally. SPP will be an accelerator and beam diagnostics test facility and it will
also serve the detector development community with its low beam current.
This paper discusses the design and construction of the RF power supply
and the RF transmission line components such as its waveguide converters
and its circulator. Additionally low and high power RF test results are presented to compare the performances 
of the locally produced components to the commercially available ones.}


\arxivnumber{1504.01576} 

\begin{document}
\maketitle
\flushbottom

\section{Introduction}
\label{sec:intro}

The SANAEM Project Prometheus (SPP) at the Turkish Atomic Energy Authority
(TAEK)'s Saraykoy Nuclear Research and Training Center (SANAEM), aims
to gain the necessary knowledge and experience to construct a proton
beamline needed for future necessities \cite{overall}. A Proof
of Principle~(PoP) accelerator with modest requirements of achieving
at least 1~MeV proton energy, with a peak beam current of few tens of $\mu$A's is under development. This PoP project has also the
challenging goal of having the design and construction of the entire
setup using local resources ranging from its ion source to the
final diagnostic station, including its RF power supply and the RF
power transmission line which are the foci of this paper. There are
also two secondary goals for this project:~1) To train the next generation
accelerator physicists and RF engineers on the job; 2) To encourage and develop skills of the local industry in the accelerator component construction.
The ion source is a pulsed 1 kW 13.56 MHz RF driven inductively coupled plasma chamber where 20 kV potential 
is used to extract 10 ms $H^{+}$ ions with a repetition frequency of 1 Hz \cite{ion_source}. The low energy proton beam's
current, profile and emittance are measured using a compact measurement station at the solenoid based LEBT line,
which was previously described elsewhere \cite{KaraKutu}. A \textquotedbl{}4-vane\textquotedbl{} Radio Frequency Quadrupole
(RFQ) operating at 352.21 MHz is used as the low beta accelerating
cavity~\cite{TepilKap}, \cite{wangler}. This operating frequency ($f$)
was selected to be compatible with similar machines in Europe and
therefore to benefit from the already available RF power supply market. The design requirements of the
accelerating cavity is given in Table~\ref{tab:SANAEM-Pop}. Other parameters such as the inter-vane voltage (V), input energy ($E_{in}$), output energy ($E_{out}$), beam current acceptance ($I$), power dissipation at maximum Q factor ($P_{d}$) and the vane length ($L$) were chosen to be adequate for a first time machine. To compensate for any possible reduction in Q factor of the manufactured cavity, a safety margin is considered and the power requirement of the RFQ is estimated as 120 kW. The accelerating cavity will be followed by a beam diagnostics section
and finally by a beam dump.

\begin{table}[htbp]
\protect\caption{SPP RFQ design parameters\label{tab:SANAEM-Pop}}
\centering{}%
\begin{tabular}{c|c}
\textbf{Parameter}  & \textbf{Value}\tabularnewline
\hline  
$E_{in}$ (MeV)  & 0.02      \tabularnewline \hline 
$E_{out}$ (MeV)  & 1.3\tabularnewline \hline 
$f$ (MHz)  & 352.21       \tabularnewline \hline 
$V$ (kV)  & 60              \tabularnewline \hline 
$I$ (mA)  & <15              \tabularnewline \hline 
$P_{d}$ (kW)  & 65.5       \tabularnewline \hline 
$L$ (m) & 1.2 \tabularnewline 
\end{tabular}
\end{table}

\section{Overview of the RF System}

\begin{figure}[htbp]
\begin{centering}
\includegraphics[width=0.7\columnwidth]{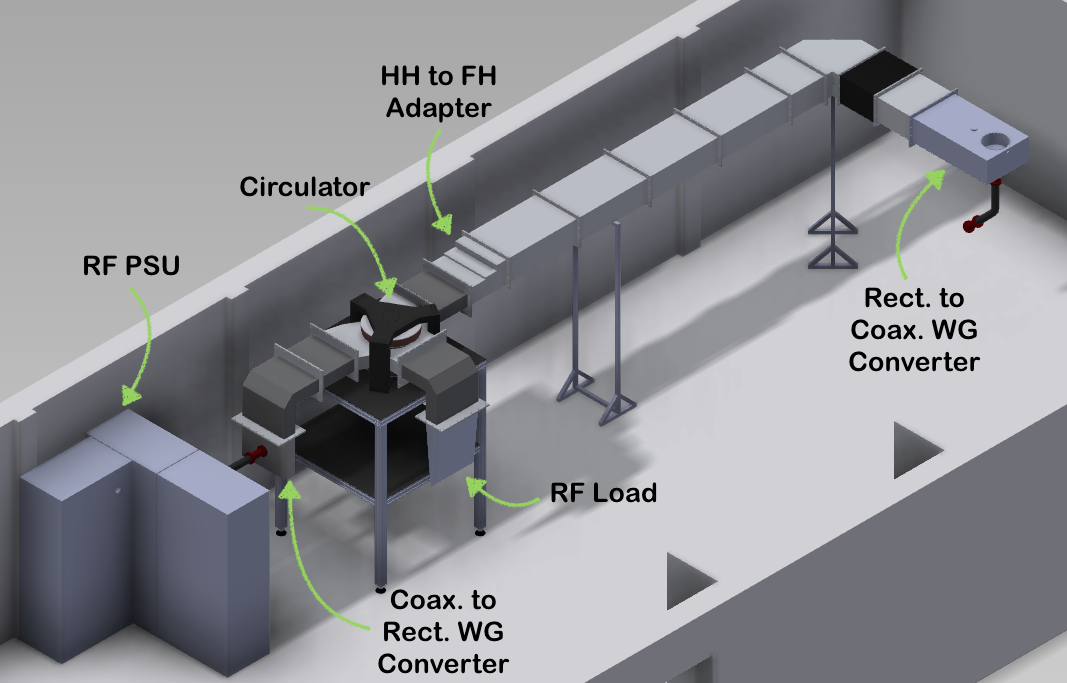}
\par\end{centering}
\protect\caption{The overview of the SPP RF transmission line. \label{fig:overview}}
\end{figure}

The overview of the RF power transmission line for the SPP project is depicted
in Fig.\ref{fig:overview}. The racks shown in lower left part of
the figure represent the RF power supply unit (PSU). The RF power supply is required
to be pulsed at a maximum duty factor of 3\% and to have a peak output
power of 120 kW at 352.21 MHz. The master oscillator's frequency precision is required to be at least 1 kHz in order to fine-tune the accelerator cavity (at this level, no LLRF system is used to follow the frequency of the cavity). A precision of 10 kW in power is considered to be enough to set the proper voltage level in the RFQ. The acceptable beam transmission limit of the RFQ cavity is set to be 70\% for the project. The cavity's inter-vane voltage should not fall below  56 kV compared to the design value of 60 kV to ensure such a transmission rate. Since 4 kV potential difference would correspond to about 16 kW difference from the PSU, its stability requirement can be written as $\pm$8 kW. Additionally, remote control capability and fast-interlocks, in the case of any failures, are other requirements for the RF PSU system.

Consequently, the PSU was designed to match
these requirements; the result being a two step amplifier, having
both solid state and vacuum tube components. The output form factor
of the PSU follows the RF amplifier cavity's output in 3 1/8" rigid coaxial
line format. A coaxial to rectangular waveguide converter has been designed to start using
a half height (HH) WR2300 rectangular waveguide leading to a HH custom designed
 circulator to protect the power supply. The RF dump is also a custom
made component, installed after a 90 degree E-bend at the third port
of the circulator. The 6 meter transmission line starts at the second
port of the circulator and is immediately converted to full height
(FH) WR2300 using a custom designed and built adapter unit. The FH
line's last components are an H-bend followed by a flexible rectangular waveguide
and finally a rectangular to coaxial waveguide converter to couple the power to the accelerating
cavity by using a magnetic loop. The rest of this manuscript is devoted
to the design, building and testing of these components.

\section{SPP RF Power Supply}
\begin{figure}[htbp]
\begin{centering}
\includegraphics[width=0.7\columnwidth]{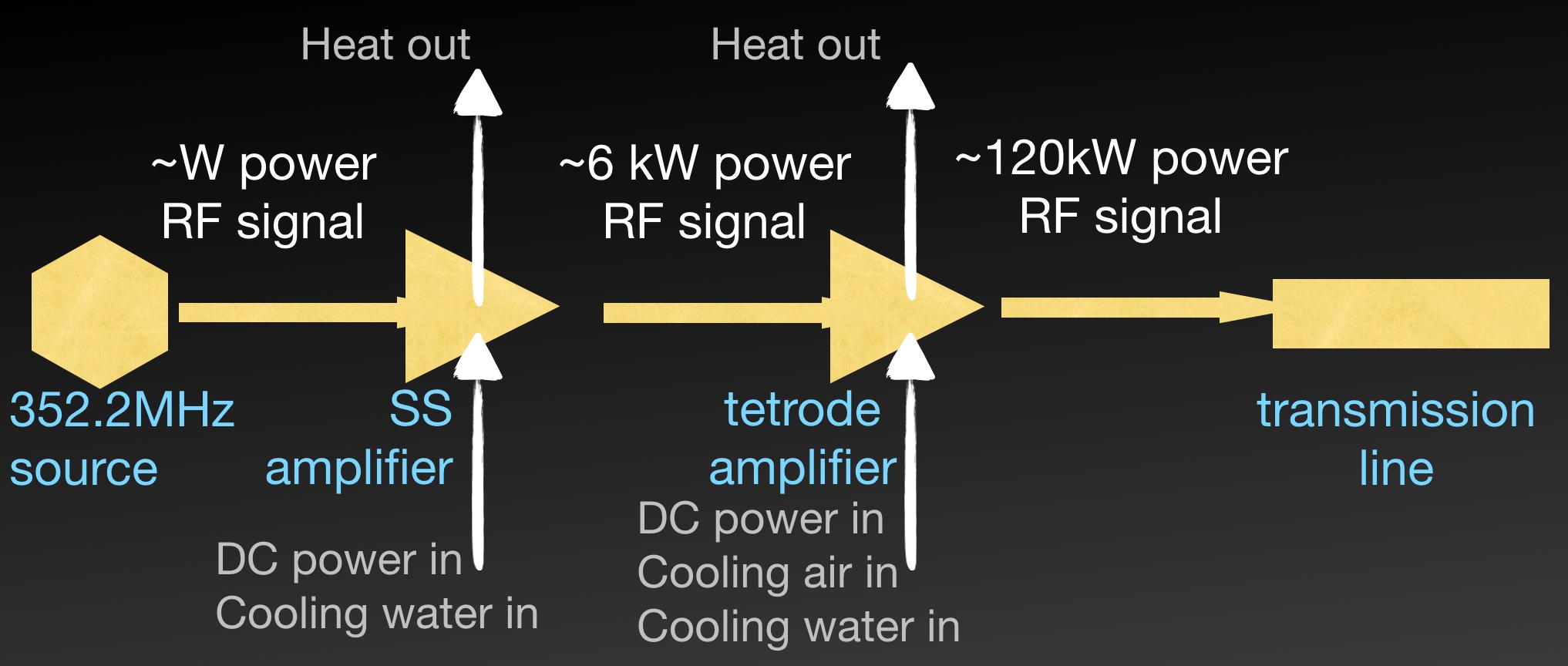}
\par\end{centering}
\protect\caption{The two stage hybrid PSU design for SPP. Note that the solid state stage is only water cooled whereas the second stage
with tetrode tube requires both water and forced air cooling.
 \label{fig:SPP-PSU-design}}
\end{figure}

The SPP power supply is required to operate at 352.21 MHz as a narrow bandwidth device, 
and to deliver about 120 kW peak power in pulsed mode at a maximum duty factor of
3\%. It is designed by the project members and manufactured by a local RF company \cite{eprom}. 
It is envisaged as a two staged amplifier, as shown in Fig.\ref{fig:SPP-PSU-design}. The first stage was implemented with a solid state amplifier and the second stage with a vacuum tube amplifier. 
The whole system was planned to allow both remote operation over IP network and local control and monitoring over the touch screen panel
on the control rack.

\begin{figure}[htbp]
\begin{centering}
\includegraphics[width=0.9\columnwidth]{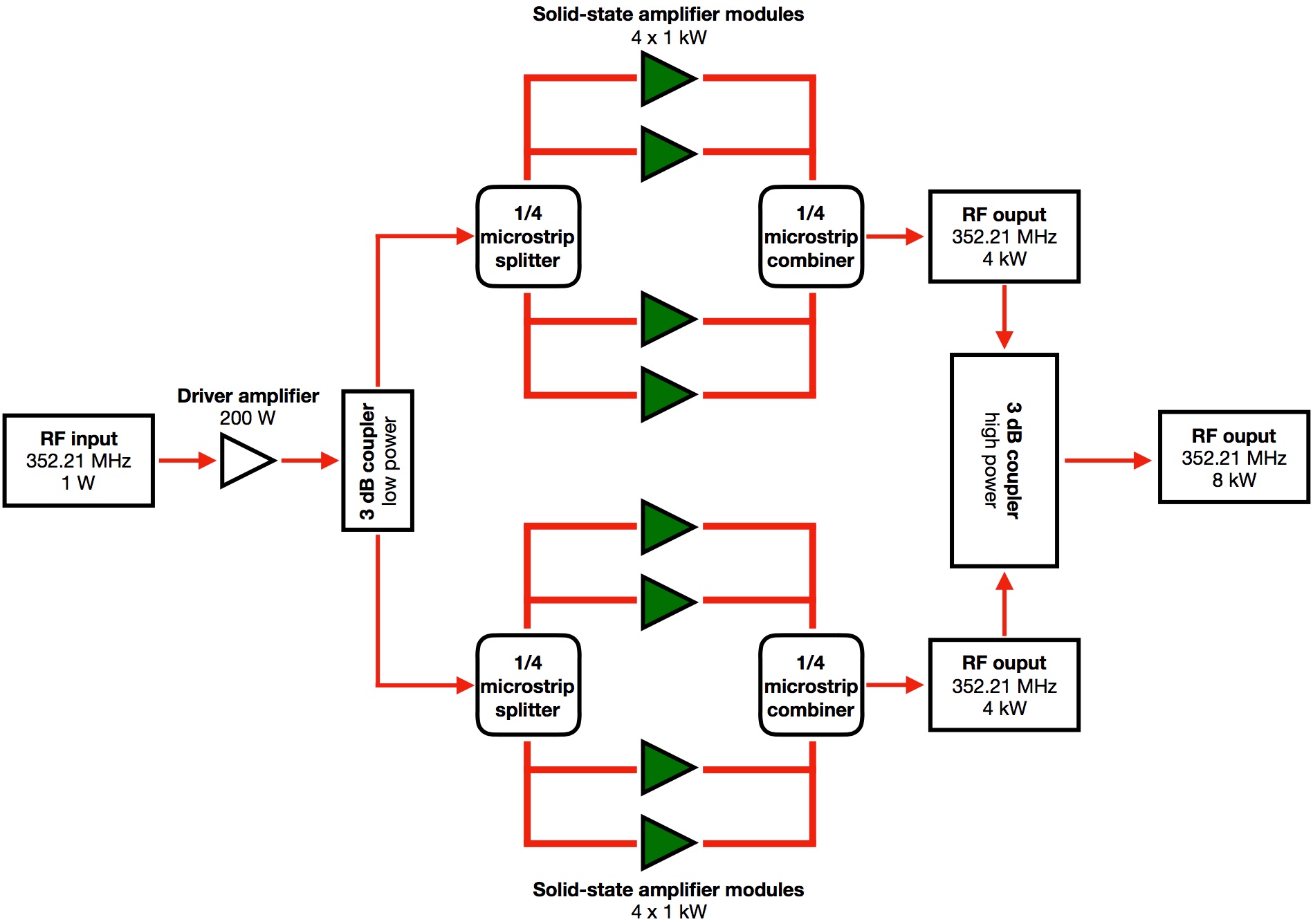}
\par\end{centering}
\protect\caption{Block diagram of the combiner module. \label{fig:combinerBlock}}
\end{figure}

The first amplification stage uses the BLF-578XR high power transistors \cite{BLF} to
achieve a combined output of 6 kW in continuous and over 8 kW in pulsed modes.  
Each solid state amplifier (SSA) circuit provides 1.1 kW per amplifier board in pulsed mode 
for pulses shorter than 500 ms. The output of stage one is, thus, obtained
by combining the outputs of eight such boards using home built combiner modules (Fig.\ref{fig:combinerBlock}).

The low power RF input of the splitter/SSA/combiner component is provided by a low noise PLL (phase-locked loop) controlled oscillator and amplified with a 1 W pre-amplifier. The driver amplifier's output power is controlled with a PLC (programmable logic controller) based gain controller to adjust the input power of the SSA modules. All error signals are evaluated with the PLC and triggering the fast-interrupts to protect the RF system, if needed. The low power 3dB coupler splits the output of the driver amplifier into two and channels these into two identical splitter/SSA/combiner boards. On these boards, a microstrip Wilkinson 1/4 splitter was designed with a teflon substrate and a 35 $\mu$m silver plated copper strip to further split the received signal and channel it into 4 identical SSA modules. Each SSA module amplifies the RF signal up 1.1 kW with the water cooled high power transistor modules. The output signals of the transistor modules are combined with a microstrip Wilkinson 1/4 combiner for both boards. Finally, the combined output of both boards are coupled up with a high power 3 dB coupler to provide 8 kW RF signal.

\begin{figure}[htbp]
\begin{centering}
\includegraphics[width=0.5\columnwidth]{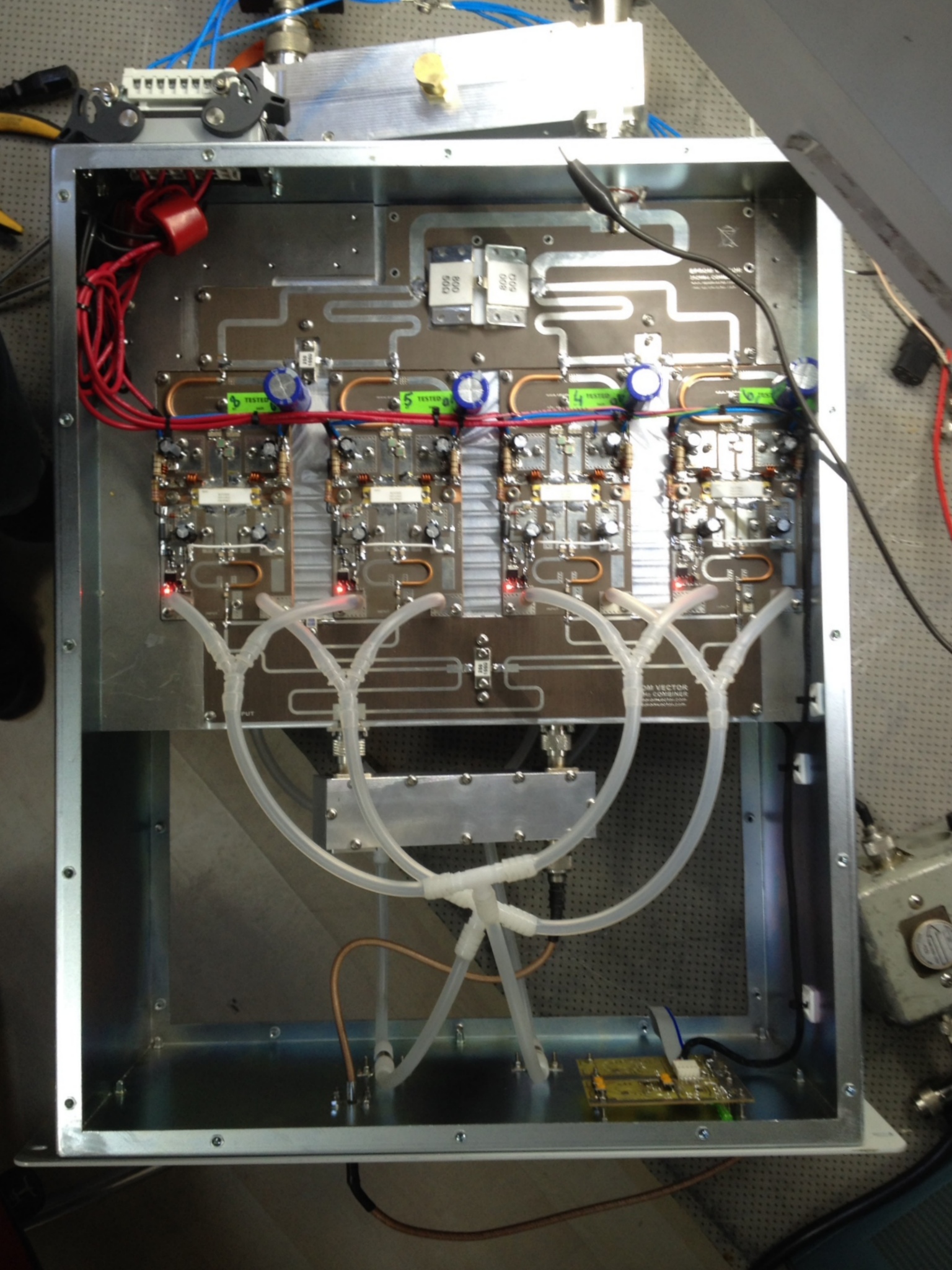}
\par\end{centering}
\protect\caption{SPP PSU solid state amplifier combiner. \label{fig:SPP-SSA-8combi}}
\end{figure}

\begin{figure}[htbp]
\begin{centering}
\includegraphics[width=0.5\columnwidth]{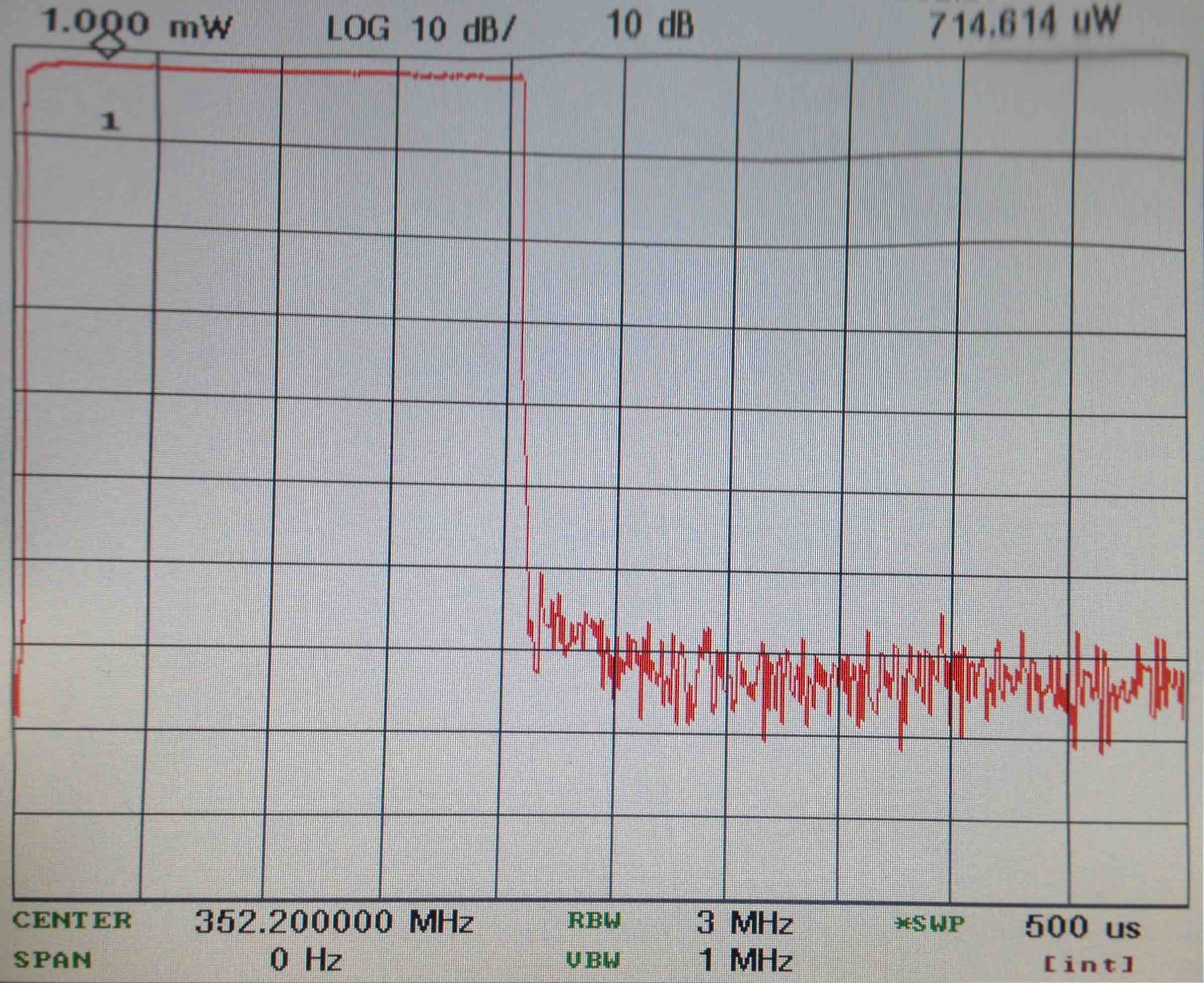}
\par\end{centering}
\protect\caption{Power test results of the pulse. \label{fig:powertest}}
\end{figure}

\begin{figure}[htbp]
\begin{centering}
\includegraphics[width=0.6\columnwidth]{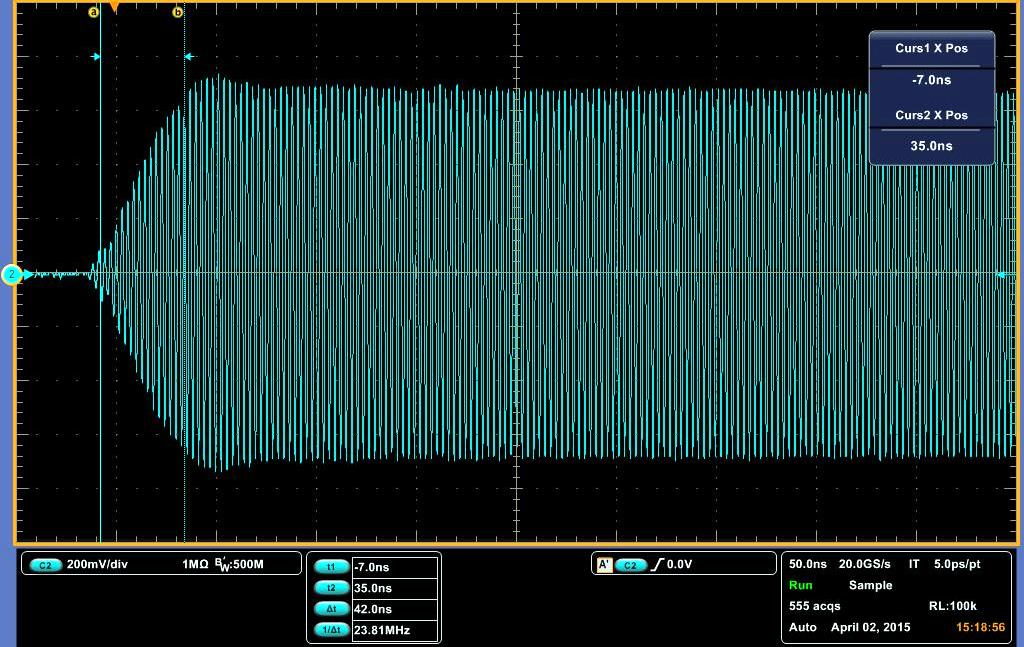}
\par\end{centering}
\protect\caption{Rise time of the RF pulse is measured as 42 ns on a dummy load. \label{fig:risetime}}
\end{figure}

The assembled combiner and a number of amplifier cards can be seen in Fig.\ref{fig:SPP-SSA-8combi}. The power and the rise time test results of the combined pulse from the SSA combiner are shown in Fig.\ref{fig:powertest} and Fig.\ref{fig:risetime}, respectively. The PSU provided about 42 ns of rise-time for an RF pulse of about 200 $\mu$s, fast enough for filling the SPP RFQ cavity.
The power tests of the SSA pulse are performed with -70 dB attenuated signal and a 7.1 kW pulse power was reached in these tests.

\begin{figure}[htbp]
\begin{centering}
\includegraphics[width=0.45\columnwidth]{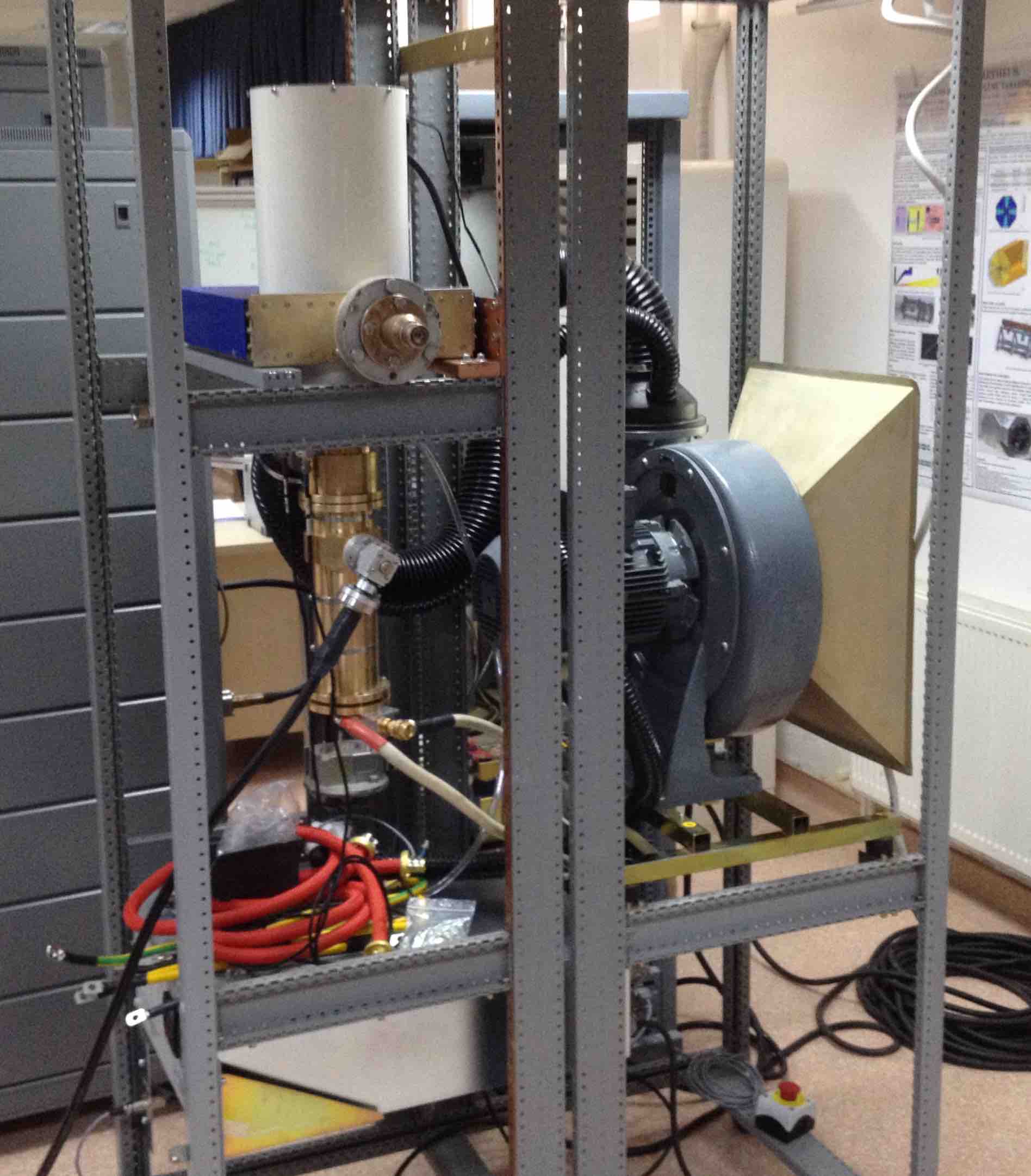}
\par\end{centering}
\protect\caption{SPP PSU Tetrode amplifier rack during assembly in the lab. Left section contains 
the RF amplifier cavity which would later house the tetrode amplifier and on the right side 
one can see the air intake for forced air cooling of the tetrode.
 \label{fig:SPP-TA-rack}}
\end{figure}

\begin{figure}[htbp]
\begin{centering}
\includegraphics[width=0.7\columnwidth]{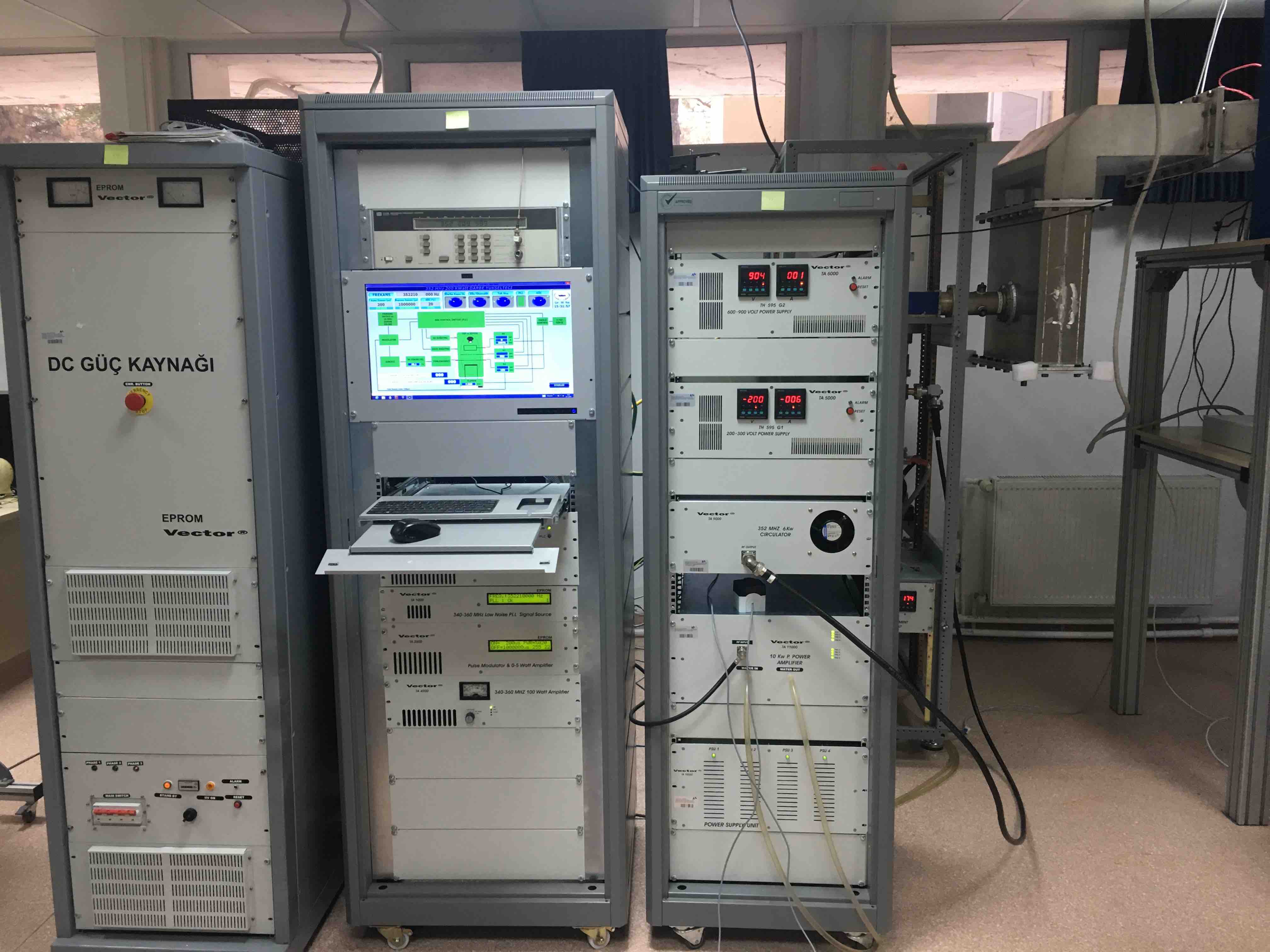}
\par\end{centering}
\protect\caption{SPP PSU in use. The left rack has the DC power supply, the central rack has the controller 
and the master signal generator, and the right rack has the power supplies for the tetrode gates, the first amplification
stage and the output to the Tetrode amplifier rack.
\label{fig:SPP-PSU_racks}}
\end{figure}

The second amplification stage uses a tetrode amplifier, both water and air cooled,
to achieve the final peak power of 120 kW. The selected tetrode tube is the 
TH595 \cite{thales} which provides up to 200 kW peak power in short pulse operation mode, 
for frequencies up to 450 MHz.  Fig.\ref{fig:SPP-TA-rack} contains a 
view of the tetrode amplifier, the amplification cavity and the forced air cooling system during assembly.
For the SPP setup, the tetrode gate voltages are adjusted as G1= -200 V and G2= 900 V, the filament voltage as 7.3 V, whereas the anode voltage is set to 13.5 kV.
The completed RF PSU including all the controls, power supplies and cooling as installed in the laboratory consists of five racks.
The DC power supply, control unit and the first amplification stage are as shown in Fig.\ref{fig:SPP-PSU_racks} during operation.

\section{SPP RF Transmission Line}
\begin{figure}[htbp]
\begin{centering}
\includegraphics[width=1\columnwidth]{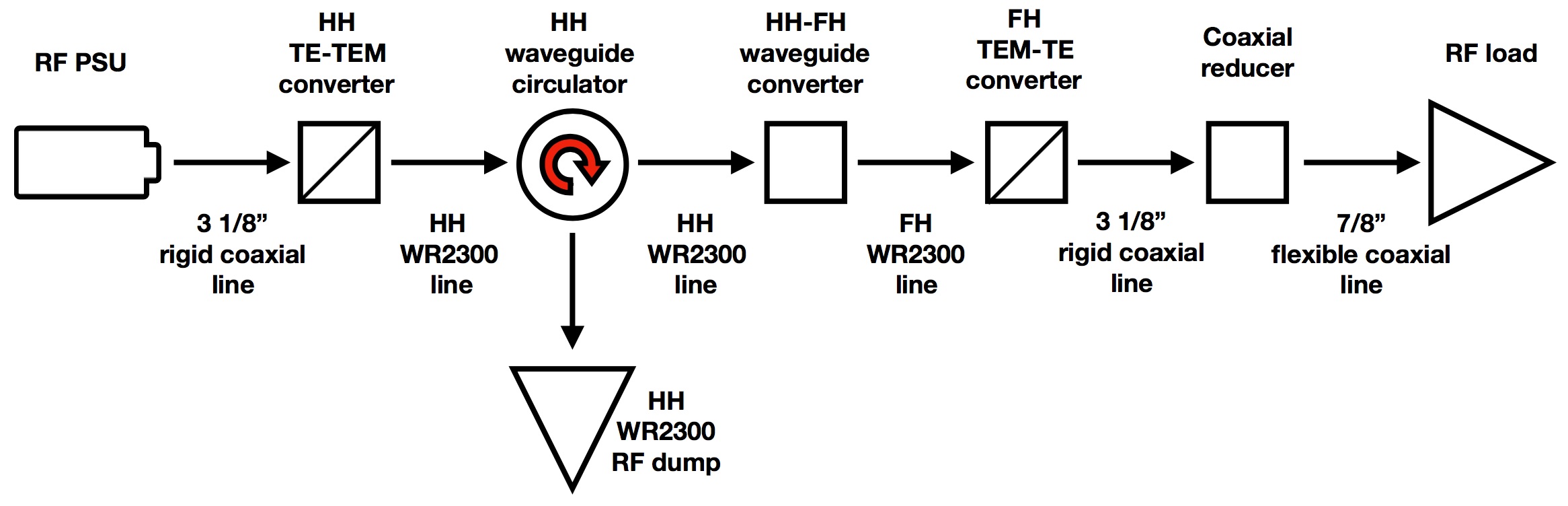}
\par\end{centering}
\protect\caption{Block diagram of the SPP RF transmission line.
\label{fig:block} }
\end{figure}

As a PoP project, SPP transmission line contains a large
variety of the available components, where 3 1/8" rigid coaxial lines, a waveguide circulator,
full height and half height rectangular waveguides (WR2300), a flexible rectangular waveguide and
their converters and adapters can be cited as examples (Fig.\ref{fig:block}). The selected rigid coaxial line 
has 50 $\Omega$ impedance and can carry a peak power of 440 kW, up
to a maximum frequency of 855 MHz \cite{rigid}. The WR2300 series rectangular waveguides
operate at a lower maximum frequency, up to 450 MHz and these can carry
a maximum power of about 700 MW. 

In the whole transmission line only one component, a flexible rectangular waveguide from Mega \cite{mega}, was purchased to ease alignment of the coaxial coupler to the RFQ. All the other RF components were first
designed using Computer Simulation Technology (CST) Microwave Studio's (MWS) Frequency Domain Solver \cite{cst} for the operating
frequency of 352.21 MHz and produced locally by the manufacturers in Middle East Industry and Trade Center 
(OSTIM) in Ankara. With the relatively modest power requirements, the SPP RF transmission line operates in normal air and doesn't have any
special isolation gas requirements. The maximum power capability requirement of the components is pulsed 120 kW with 0.01\% duty factor and 1 Hz repetition frequency.
Upon delivery, reflection ($S11$) and
transmission ($S21$) characteristics of the components were measured at the operating
frequency using low power output from a vector network analyzer (VNA) \cite{network-analyzer}. Lastly, the high power tests were performed with the RF PSU and the whole RF transmission line.

\subsection{RF Dump, Bends and HH to FH Transition\label{Dump}}

\begin{figure}[htbp]
\begin{centering}
\includegraphics[width=0.35\columnwidth]{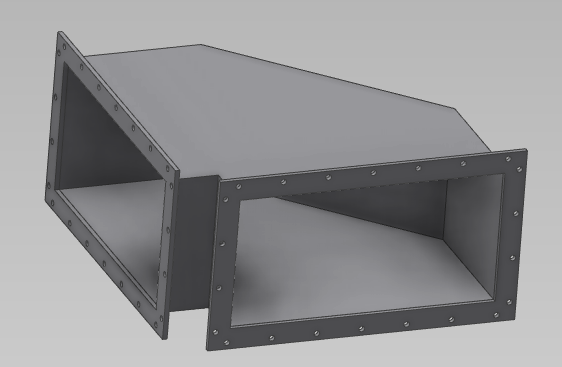}
\includegraphics[width=0.28\columnwidth]{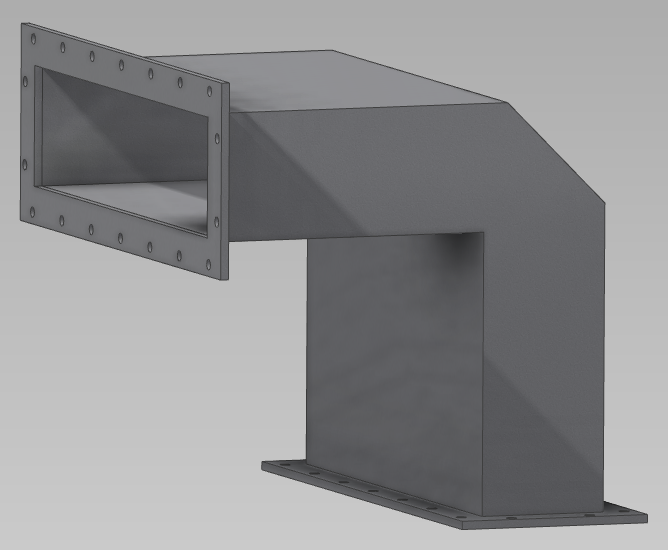}
\par\end{centering}
\protect\caption{ Full height H-bend(left) and half height E-bend (right) designs for SPP. \label{fig:Bends} }
\end{figure}

\begin{figure}[htbp]
\begin{centering}
\includegraphics[width=0.63\columnwidth]{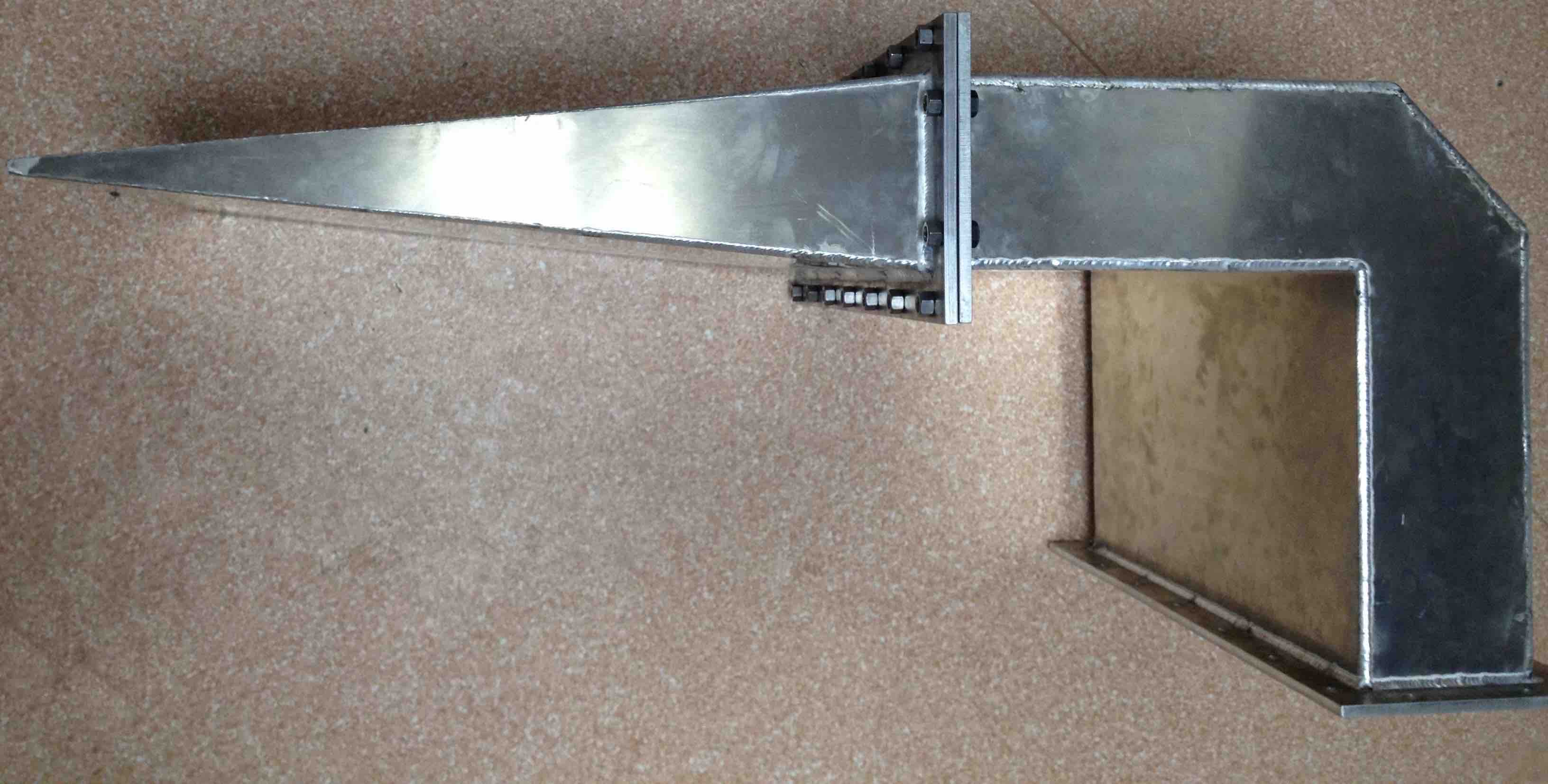}
\par\end{centering}
\protect\caption{Half height E-bend attached to RF dump.
\label{fig:ebend-dump} }
\end{figure}

The half height RF dump is made from a triangularly ending waveguide
loaded with silicon carbide. Its length is about 50 cm and it can
be air cooled for high power operations. The two E-bends are to be used
with the circulator setup, whereas the H-bend is for the section of the transmission line
just before the RFQ. Therefore the bends are designed as HH and FH modules as shown in Fig.\ref{fig:Bends}. These components are optimized to perform below $-85$ dB for $S11$  (with PEC assumption) and produced as seen in Fig.\ref{fig:ebend-dump} and Fig.\ref{fig:The-whole-s21}, respectively.

\begin{figure}[htbp]
\begin{centering}
\includegraphics[width=0.4\columnwidth]{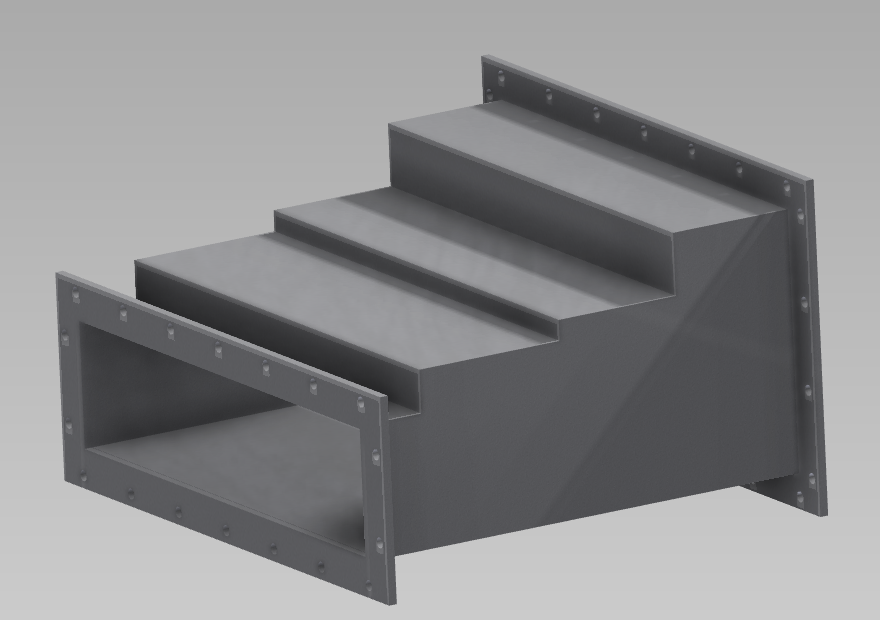}
\includegraphics[width=0.375\columnwidth]{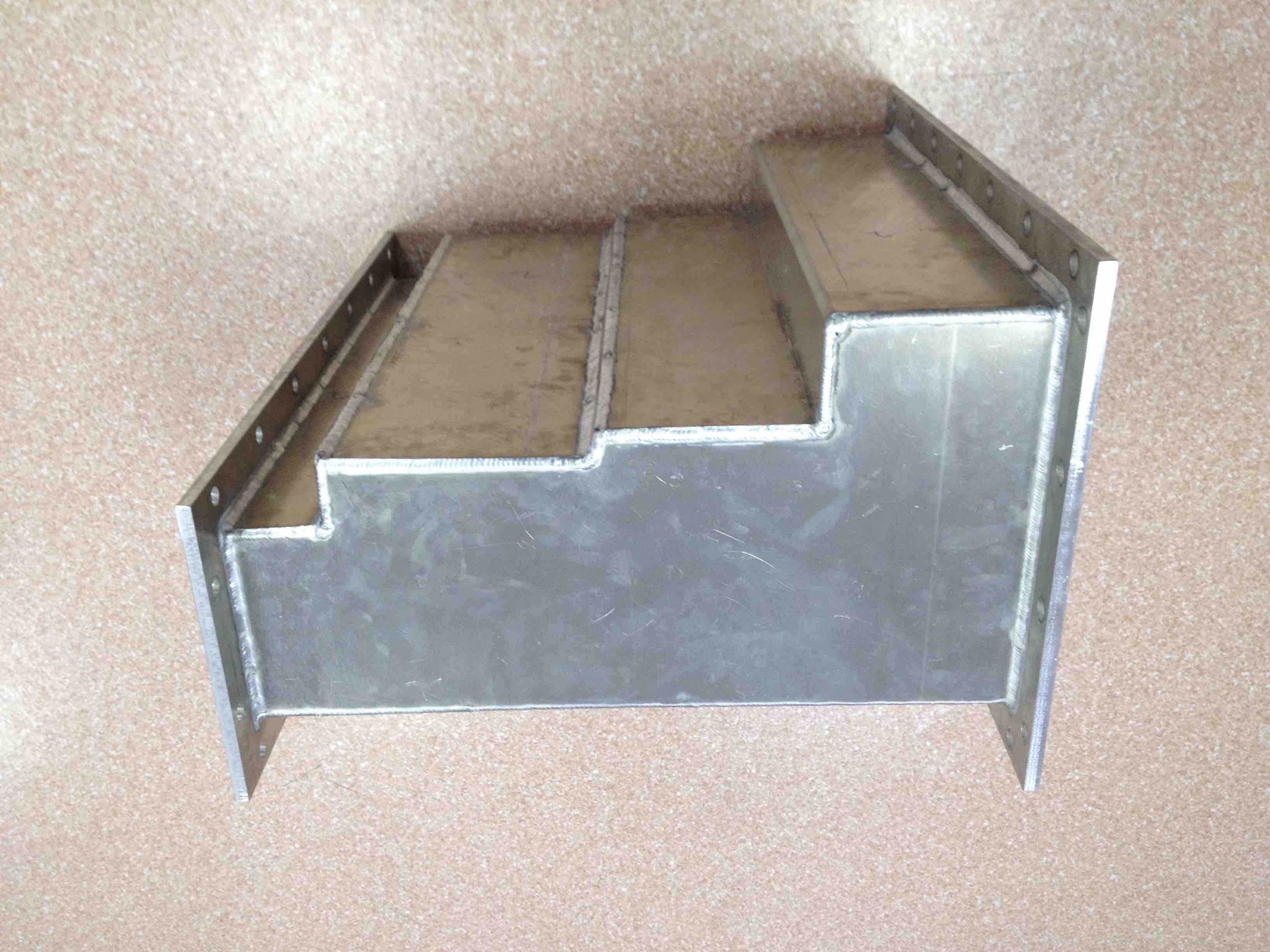}
\par\end{centering}
\protect\caption{\emph{Left}: HH to FH transition design. \emph{Right}: The finished product.
\label{fig:HHtoFH} }
\end{figure}

Fig.\ref{fig:ebend-dump} contains the manufactured and assembled HH E-bend attached to the RF dump to tune the coaxial-rectangular waveguide converters just before integration to the RF transmission line (Fig.\ref{fig:The-S11-setup}). As the RF transmission line contains both FH and HH
sections, a transition module was necessary. The design of the HH to FH transition was performed by
using CST MWS while optimizing the total length, the transmission efficiency and the producibility (Fig.\ref{fig:HHtoFH}). 
Considering these factors, the $S11$ of the HH to FH transition is optimized to $-61$ dB, similarly with PEC assumption.

\subsection{Measurements with Coaxial-Rectangular Waveguide Converters, Reducers and Other Parts}

\begin{figure}[htbp]
\begin{centering}
\includegraphics[width=0.7\columnwidth]{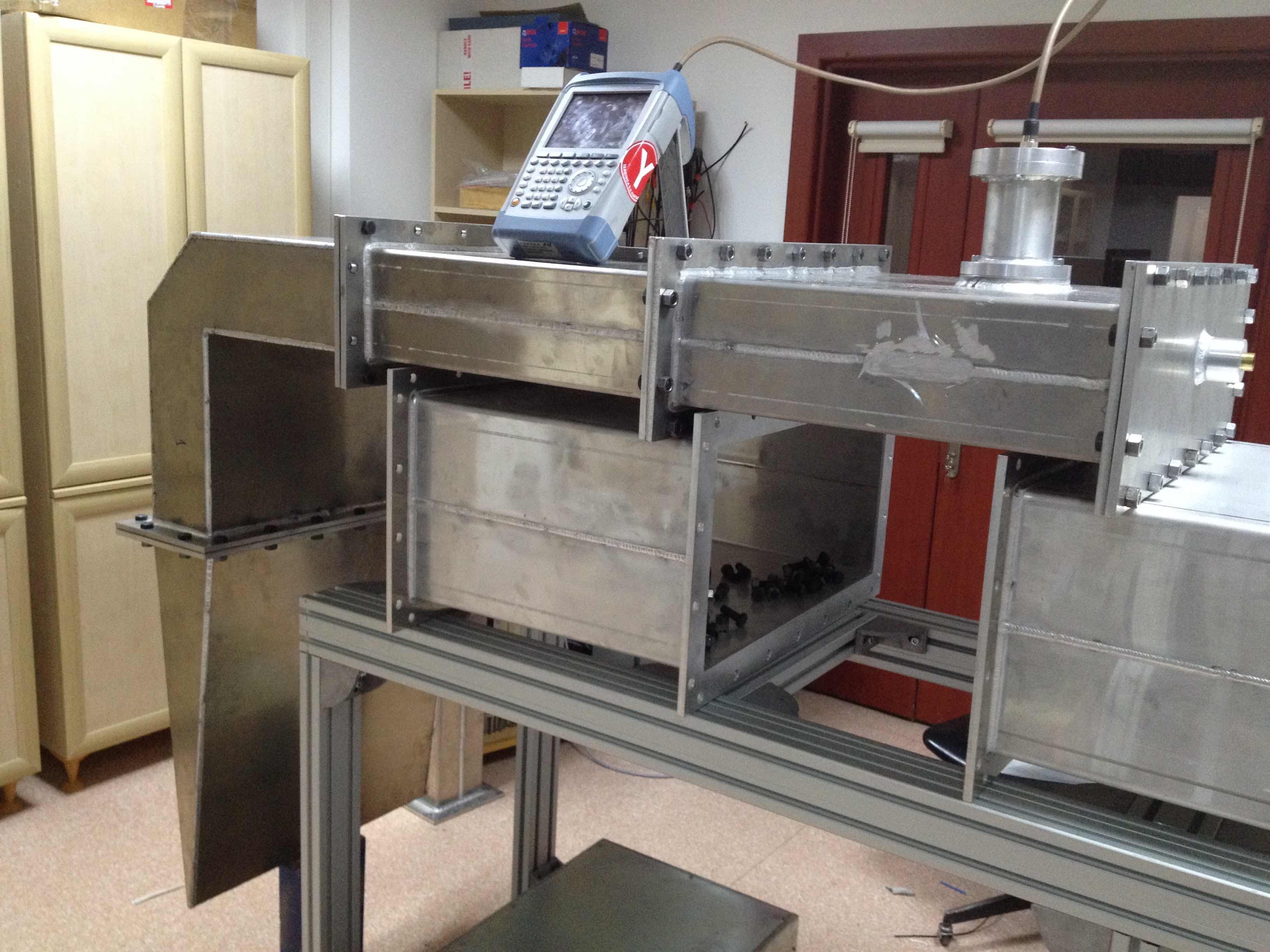}
\par\end{centering}
\protect\caption{The $S11$ measurement setup for HH TEM to TE converter. \label{fig:The-S11-setup} }
\end{figure}

\begin{figure}[htbp]
\begin{centering}
\includegraphics[width=0.5\columnwidth]{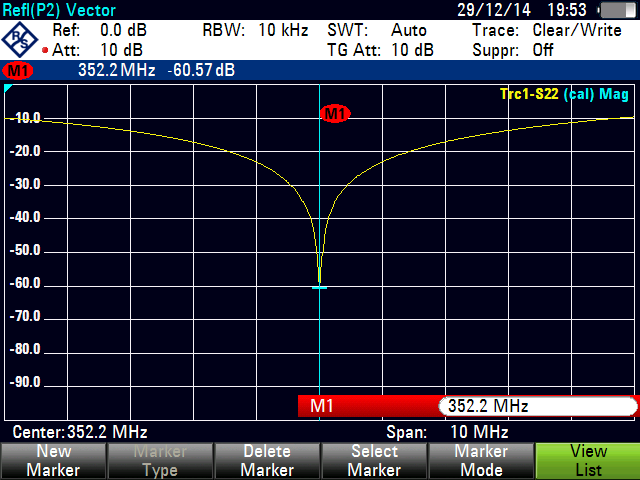}
\par\end{centering}
\vspace{0.2cm}
\begin{centering}
\includegraphics[width=0.5\columnwidth]{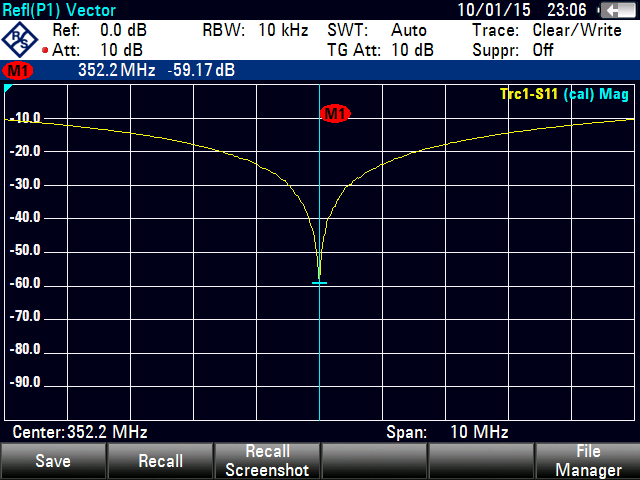}
\par\end{centering}
\protect\caption{\emph{Top}: $S11$ for TEM to TE conversion. \emph{Bottom}: $S11$ for TE
to TEM conversion. Both measurements are made by using a low power RF signal from the VNA.
 \label{fig:S11-olcumleri}}
\end{figure}

The $S11$ measurements were performed using the VNA, which provided a low power RF signal to the HH coaxial to rectangular (TEM to TE) waveguide converter with a home-built 3 1/8" to N type reducer. The 50 cm converter
was attached to a 50 cm HH WR2300 waveguide, the 90 degree E-bend
and finally to the RF dump as seen in Fig. \ref{fig:The-S11-setup}.
The low power pulse of the VNA, eventually
absorbed by the RF dump, was used to tune the converters. Before the production phase, the $S11$ parameters of the reducer and the TEM to TE waveguide converter are optimized to $-69$ dB and $-62$ dB, respectively, with the above mentioned simulation setup. The maximum CW power capability of the TEM to TE waveguide converter was found as 400 kW without any cooling.
Fig.\ref{fig:S11-olcumleri}
contains the $S11$ measurement results for the previously described
setup. After tuning the TEM to TE waveguide converter, at the operating frequency of 352.21 MHz, the attained
$S11$ value was $-60$ dB.

 As previously discussed, the RF transmission line also requires a FH rectangular to coaxial (TE to TEM) waveguide converter. For the TE to TEM waveguide converter the $S11$ parameter was optimized to $-63$ dB with computer simulations. The CW power handling capacity of the TE to TEM waveguide converter was calculated as 275 kW without any cooling. After the production of the TE to TEM waveguide converter the second measurement employs
the same setup, except the HH TEM to TE waveguide converter was replaced by
the mentioned HH to FH adapter and the FH TE to TEM waveguide converter. 
The $S11$ for this case read as $-59$ dB after tuning, which can be
found in the same figure, second screenshot. The HH TEM to TE and FH TE to TEM waveguide converters were designed as narrow-band devices due to the RFQ cavity's requirements. For a comparison, an RF-lambda \cite{rflambda} product RFWA2300 has 1.3 VSWR in a bandwidth of 370 MHz and our converters have bandwidths of 4 MHz. By means of this narrow-bandwidth, designed converters have a sharper S11 and are less lossy at the operating frequency of the cavity.

\begin{figure}[htbp]
\begin{centering}
\includegraphics[width=0.6\columnwidth]{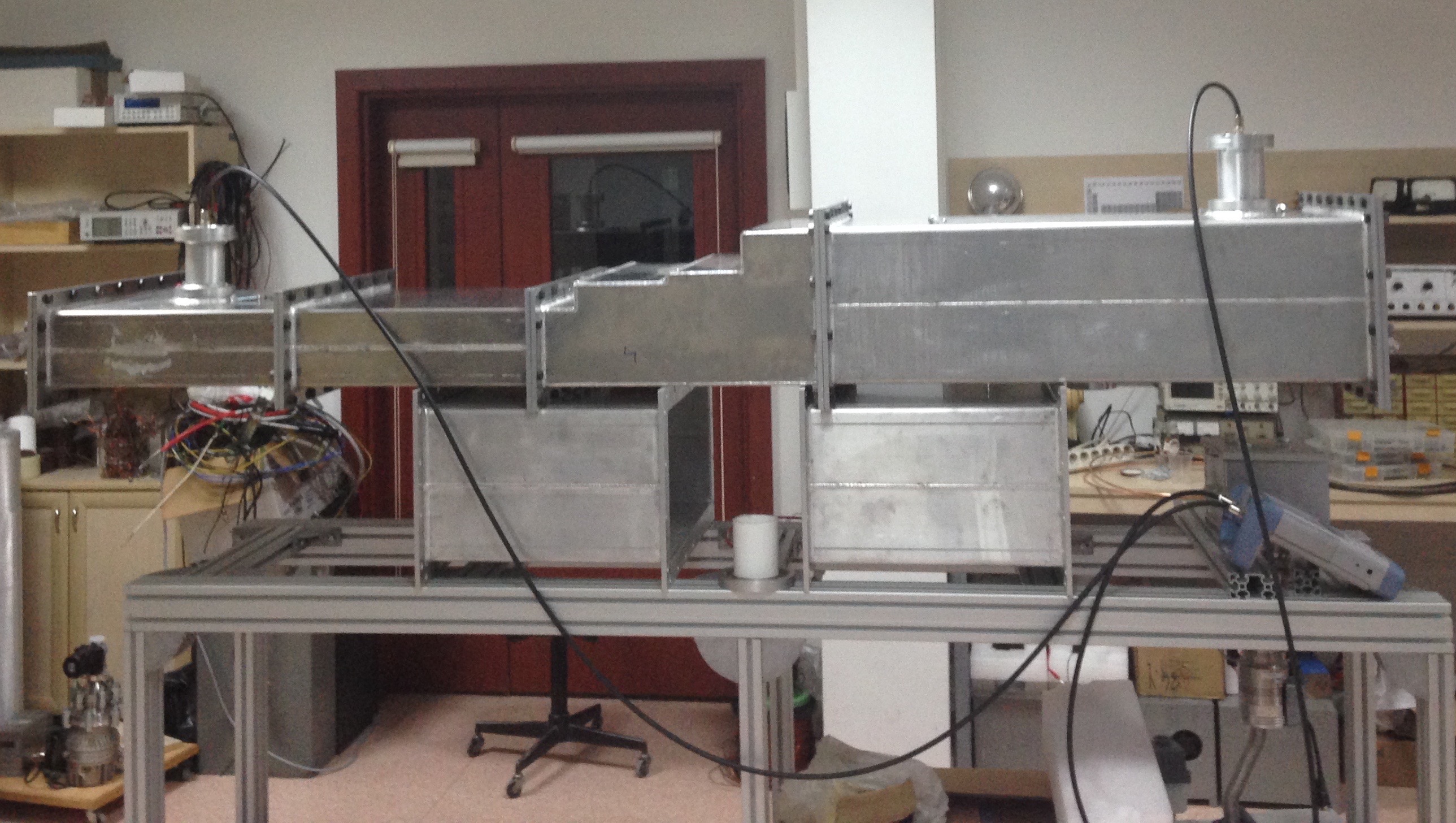}
\par\end{centering}

\protect\caption{The $S21$ measurement setup of the converters. \label{fig:The-s21}}

\end{figure}

Insertion loss measurement setup is given in Fig.\ref{fig:The-s21}. Basically a low power pulse is emitted by the
VNA which is converted to TE mode and propagated from
left to right through HH waveguide, HH to FH adapter and finally
FH TE to TEM waveguide converter, all being previously discussed. The pulse arriving
to the right side is received by the same VNA which
can calculate the $S21$ parameter. Its untuned measured value for the operating
frequency of 352.21 MHz is $-0.17$ dB.

\begin{figure}[htbp]
\begin{centering}
\includegraphics[width=0.5\columnwidth]{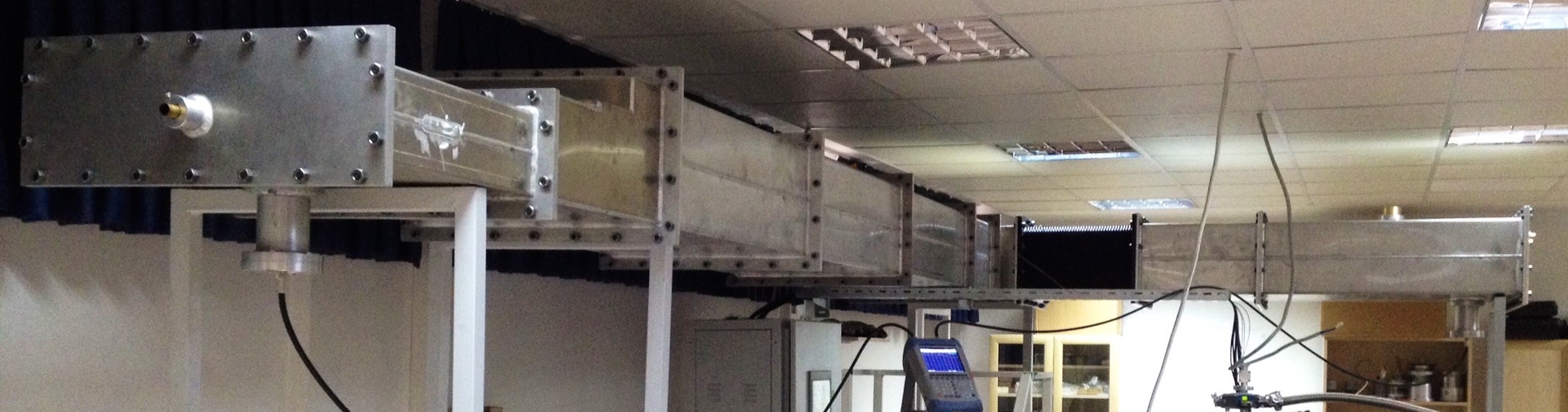}
\par\end{centering}
\vspace{0.2cm}
\begin{centering}
\includegraphics[width=0.5\columnwidth]{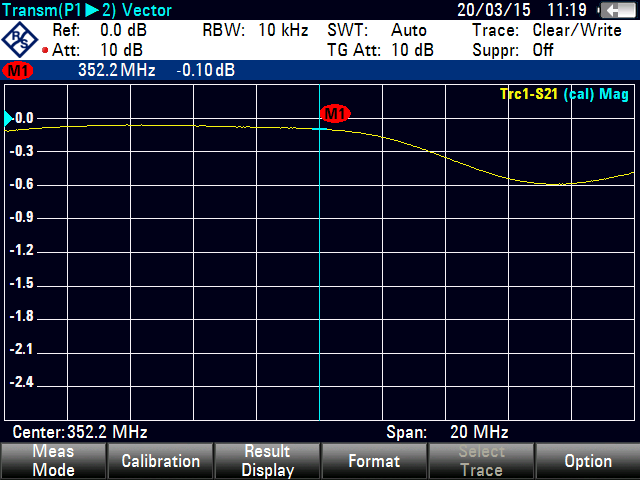}
\par\end{centering}
\protect\caption{\emph{Top}: The $S21$ measurement setup of the transmission line. A low power RF pulse from the VNA goes from left to right and finally back the VNA.
\emph{Bottom}: The $S21$ results of the RF transmission line. \label{fig:The-whole-s21}}
\end{figure}

Following the low power tests, the most of the waveguide components of the RF transmission line were assembled as in Fig.\ref{fig:The-whole-s21}. To measure the $S21$ parameter of the whole transmission line, the same procedure was followed. For the SPP operating frequency, $S21$ of the entire line was measured as $-0.1$ dB, corresponding to about 2\% power loss at the whole chain, including both of the reducers which are the main source of the power loss in the low power tests (Fig.\ref{fig:The-whole-s21}). 

\begin{figure}[htbp]
\begin{centering}
\includegraphics[width=0.6\columnwidth]{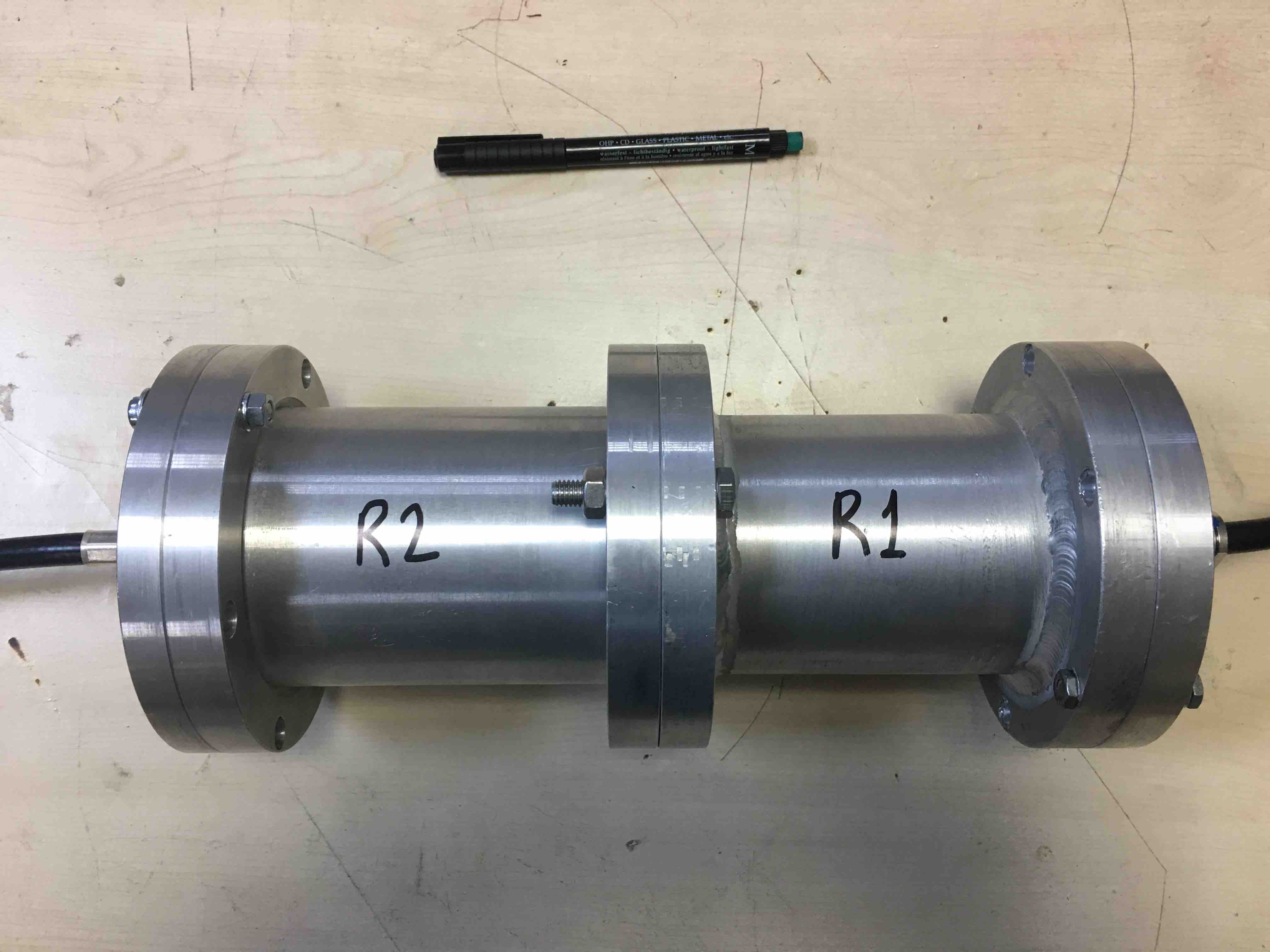}
\par\end{centering}
\protect\caption{The $S21$ measurement setup for the reducers. \label{fig:reducer} }
\end{figure}

The reducers were also tested with the VNA pulse and an insertion loss of -0.04 dB was recorded (Fig.\ref{fig:reducer}).

\subsection{Waveguide Circulator}

\begin{figure}[htbp]
\begin{center}
\includegraphics[scale=0.5]{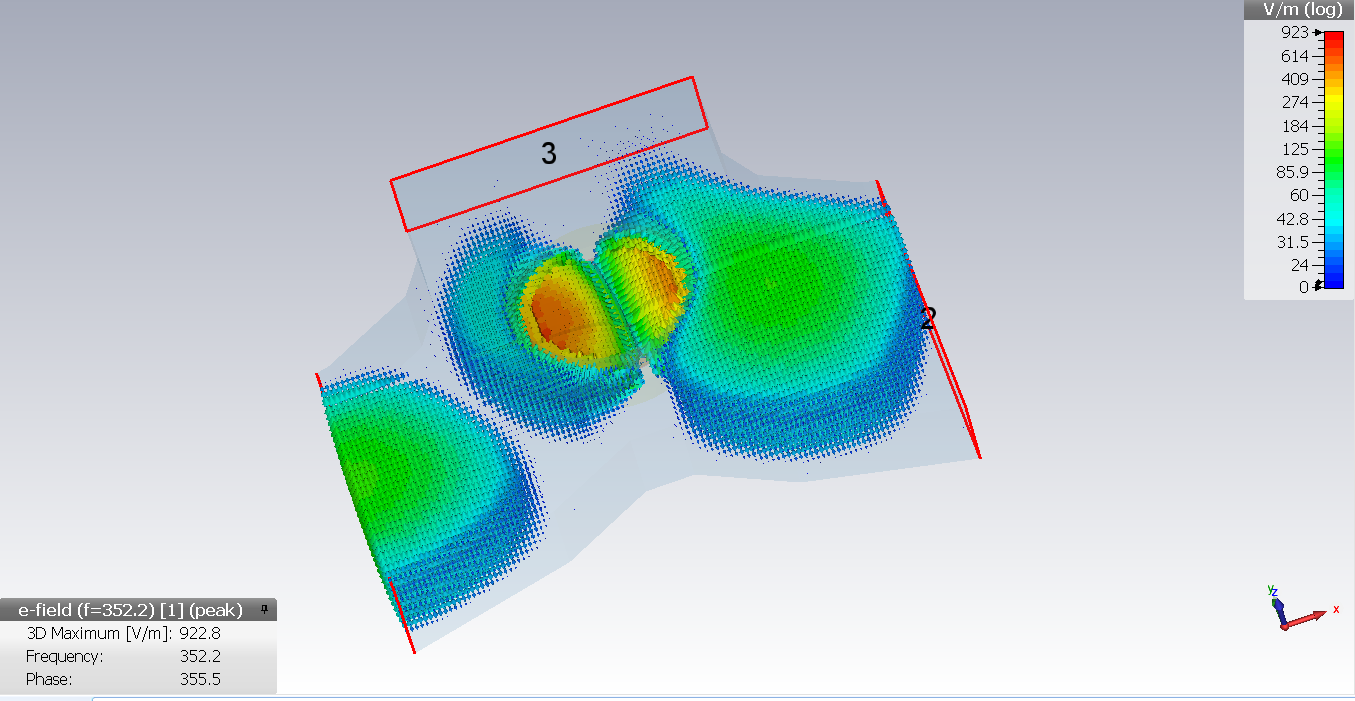}
\caption{Electric field inside the circulator for 352.21 MHz. \label{circ_elec}}
\end{center}
\end{figure}

A circulator is a non-reciprocal passive RF transformer, used to deliver the RF power to a target device, e.g. an accelerator cavity, while protecting the RF power supply with an RF load in the event of an impedance mismatch between the supply and the load. An H-plane waveguide junction circulator,  to be operated at the frequency of 352.21 MHz, has been designed using CST MWS \cite{cst} for the SPP project. This RF device is intended to deliver the RF power generated by the RF power supply unit, with minimal loss, while protecting the power supply against any power reflections. The maximum power handling requirement for the project is about 120 kW which will be supplied from port-1. The  circulator is designed to transfer this power to port-2, as seen in Fig.\ref{circ_elec}. 

That port is connected to the RF transmission line,  delivering RF power to the RFQ  by the means of waveguides and rigid coaxial cables, discussed previously. The air cooled RF dump that was discussed in the previous section \ref{Dump} is connected to port-3 to absorb any reflections from port-2.

\begin{figure}[htbp]
\begin{center}
\includegraphics[scale=0.5]{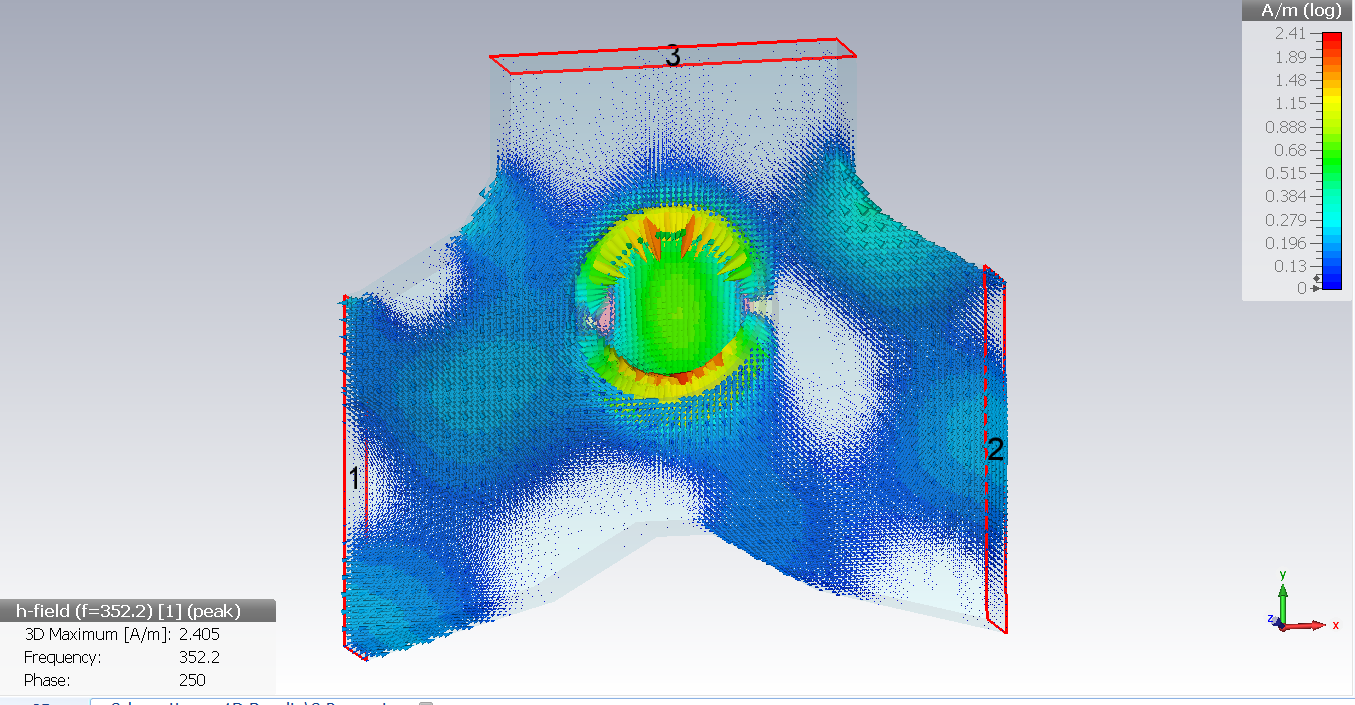}
\caption{Magnetic field inside the circulator for 352.21 MHz. \label{circ_mag1}}
\end{center}
\end{figure}

The designed circulator consists of two aluminum disks and two disks of ferrites which are attached inside the HH WR2300 waveguide. Following CST MWS simulations, the ferrite material is determined as NG-1600 from the catalogue of Magnetics Group \cite{ferrite-satici}. NG-1600 has a saturation magnetization $4\pi M_s$ at 1600 Gauss, its maximum line width is 10 Oersted, and its dielectric constant is 14.6 while the loss tangent is below $2\times 10^{-4}$, where all values are given by the supplier at 9.4 GHz. The Curie temperature is given as 220$^\circ$C which was considered in the thermal simulations. The necessary setup to keep the ferrites at room temperature during the operation was made available via water cooling. However due to low RF duty factor (0.01\%), so far no such need has arisen.
The relation between the saturation magnetic field of the ferrite $\vec{M_s}$ and applied field $\vec{H}$ both having time dependence of $\omega$  can be formulated as below:
\begin{equation}
\begin{bmatrix}M_{x}\\
M_{y}\\
M_{z}
\end{bmatrix}=\begin{bmatrix}X_{xx} & X_{xy} & 0\\
X_{yx} & X_{yy} & 0\\
0 & 0 & 0
\end{bmatrix}\begin{bmatrix}
H_{x}\\
H_{y}\\
H_{z}
\end{bmatrix},
\end{equation}
where elements of $X$ correspond to; 
\begin{equation}
X_{xx}=\dfrac{\omega_{m}\omega_{0}}{(\omega_{m}^{2}-\omega^{2})}=X_{yy},
\end{equation}
\begin{equation}
X_{xy}=\dfrac{j\omega\omega_{m}}{(\omega_{m}^{2}-\omega^{2})}=-X_{yx}.
\end{equation}
Thus, magnetic field becomes; 
\begin{equation}
\vec{B}=\mu_{0}\left(\begin{bmatrix}1 & 0 & 0\\
0 & 1 & 0\\
0 & 0 & 1
\end{bmatrix}+\begin{bmatrix}X_{xx} & X_{xy} & 0\\
X_{yx} & X_{yy} & 0\\
0 & 0 & 0
\end{bmatrix}\right)\vec{H}=[\mu]\vec{H},
\end{equation}
where the Larmor and magnetization frequencies are respectively; 
\begin{equation}
\omega_{0}=\mu_{o}\gamma H_{0},
\end{equation}
\begin{equation}
\omega_{m}=\mu_{o}\gamma M_{s},
\end{equation}
where, $\mu_0$ is vacuum permeability and $\gamma$ is the gyromagnetic ratio. The permeability of the ferrite $\mu$ has been discussed in detail by Wu and Du \cite{Wu}. Comparing these two frequencies, it can be seen that this circulator is working above the resonant frequency. An advantage of this operation regime is to have lower field losses as pointed out by Bosma \cite{Bosma}. The fields inside the circulator are so-called the acceleration mode which is TM$_{10}$ (or in European notation E$_{10}$) and the view of the magnetic field can be seen from Fig.\ref{circ_mag1}, while the electric field is already presented in Fig.\ref{circ_elec}.

\begin{figure}[htbp]
\centering{}\includegraphics[width=0.8\columnwidth]{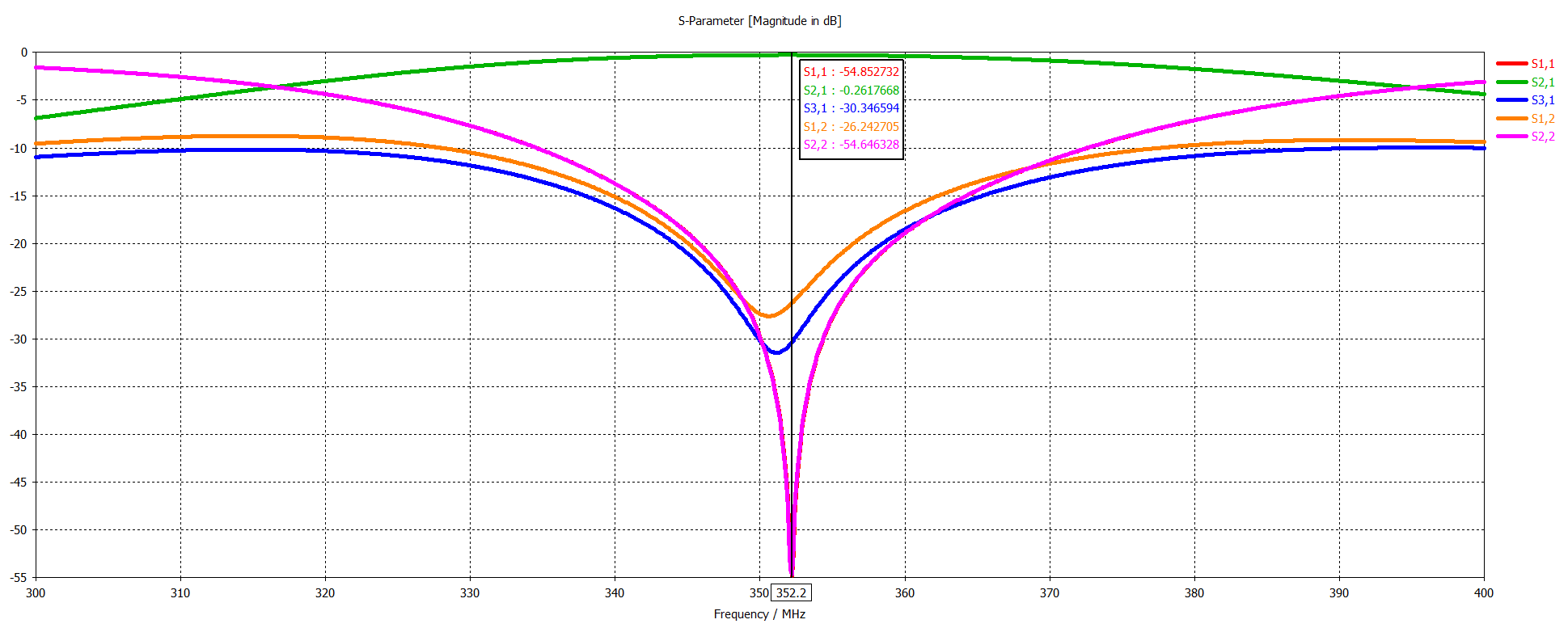} \protect\caption{Performance of the circulator design at 352.2 MHz. \label{fig:circulator_results}}
\end{figure}

The performance of the designed circulator, obtained after lengthy optimization studies, can be seen in Fig.\ref{fig:circulator_results}. At the SPP operating frequency of 352.21 MHz the return loss ($S11$), insertion loss ($S21$) and the isolation loss ($S31$) are computed as -55 dB, -0.26 dB and -30 dB, respectively. VSWR is calculated to be about 1.005:1. The power handling capacity of the circulator is about 30 kW CW without needing any water cooling. As can be seen in Fig.\ref{fig:circulator_results} the bandwidth of the designed circulator is 12 MHz for a maximum VSWR of 1.2. For a comparison, an RF-lambda product RFWC2300A having 1.2 VSWR in a bandwidth of 17 MHz, has -0.3 dB maximum insertion loss and -20 dB minimum isolation loss. Therefore the performance of this locally designed and manufactured circulator is comparable (-0.26 dB vs -0.3 dB and  -30 dB vs  -20 dB) to an industrial one. 

Finally, the biasing magnetic field of the circulator is supplied by a locally designed and built dipole electromagnet to fine tune of the ferrites. The dipole is able to produce an external magnetic field of 2.3 kG to the ferrites without any cooling requirements.

\subsubsection{Assembly, Installation and Tests}
\begin{figure}[htbp]
\centering{}\includegraphics[width=0.7\columnwidth]{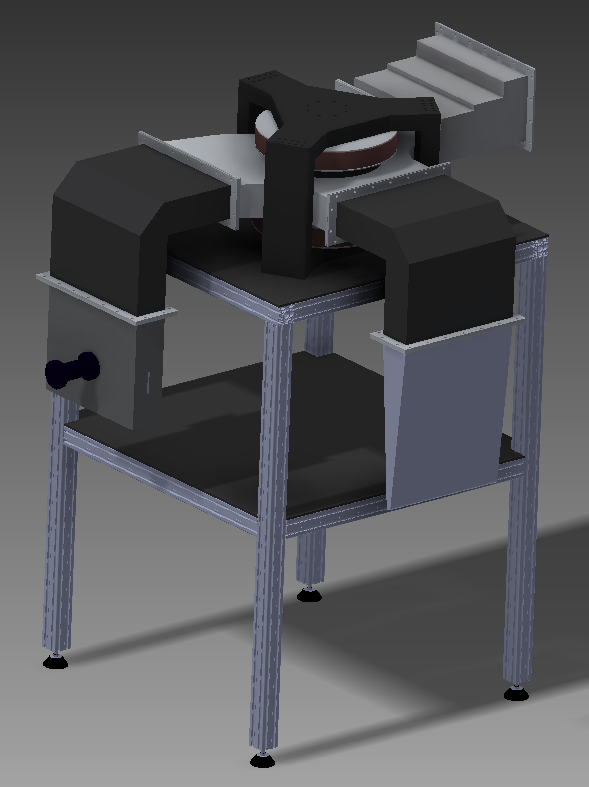} \protect\caption{The SPP HH waveguide circulator together with its
TEM to TE waveguide converter on port-1 (left side), its RF dump on port-3 (lower right) and its RF transmission line connection on port-2 (upper right).    \label{fig:circulator_drawing}}
\end{figure}

Most of the circulator components have been bought or produced locally, except the two ferrite disks purchased from TSI ceramics \cite{ferrite-satici}.
The non-magnetic components were constructed from 6061 series aluminum and attached to a support frame made from
2 cm thick iron layers that also served as a return yoke for the magnetic field. 
For field tunability, the biasing magnets were made solely from two solenoids forming a dipole. 
The coils of the dipole are designed to carry a current of about ~17 A each. 
The total weight of the circulator including the coils and yokes is about 1 Ton.
The 3D drawing of the whole circulator setup is given in Fig.\ref{fig:circulator_drawing}.

\begin{figure}[htbp]
\centering{}\includegraphics[width=0.5\columnwidth]{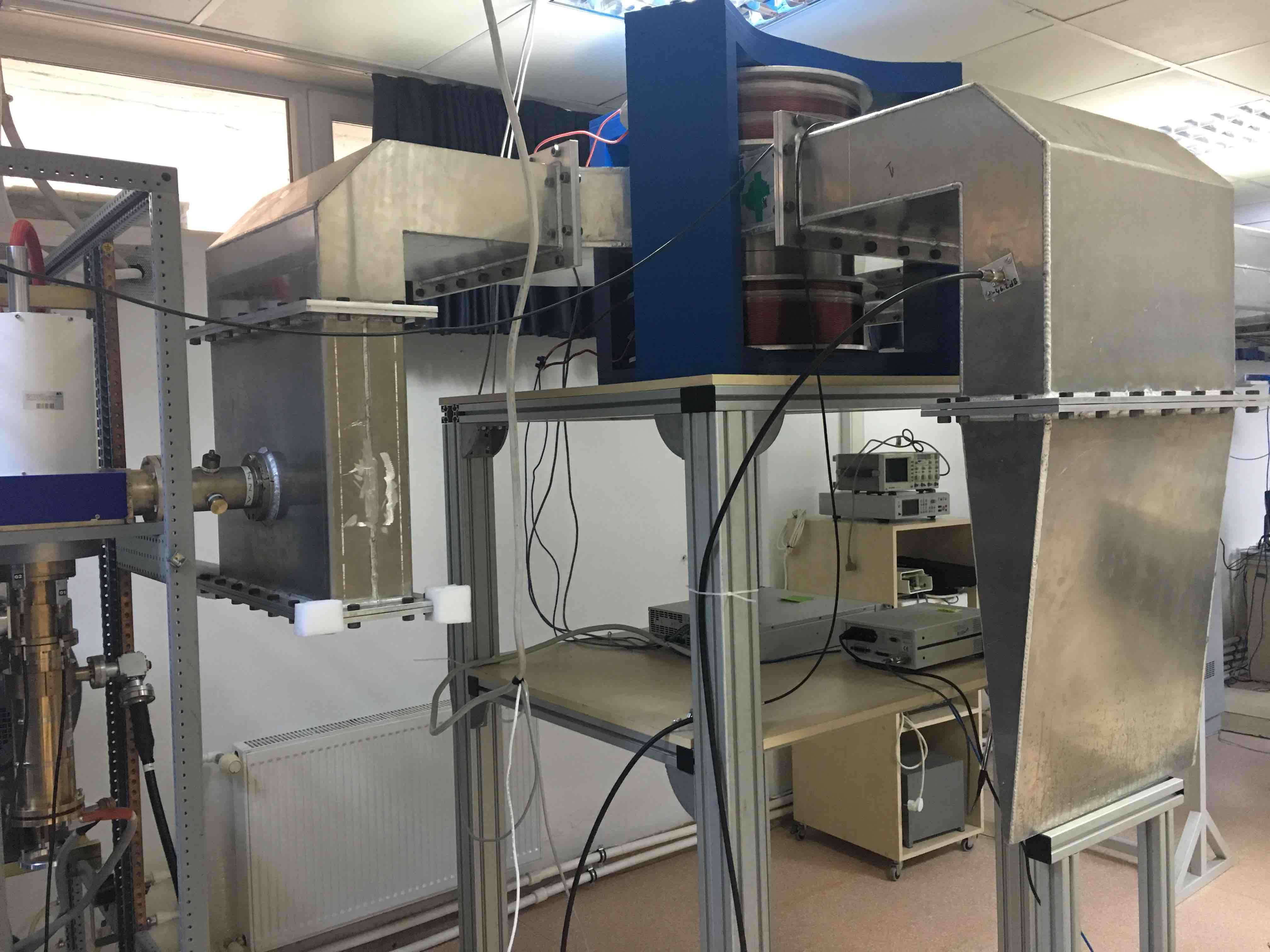} \protect\caption{352.21 MHz waveguide circulator as it is installed in the RF transmission line.
Note that the connection to the RF PSU (port-1 one left side) was made with a 3 1/8" rigid coaxial line. The electromagnet, PSU and the gauss-meter are right below the circulator.
  \label{fig:circulator_picture}}
\end{figure}

The individual components were assembled in the laboratory by the team members and installed in the RF transmission line as seen in Fig.\ref{fig:circulator_picture}.
The biasing field estimated in CST as 1910 G has been optimized as 1880 G after the installation using a VNA.

\begin{figure}[htbp]
\centering{}\includegraphics[width=0.5\columnwidth]{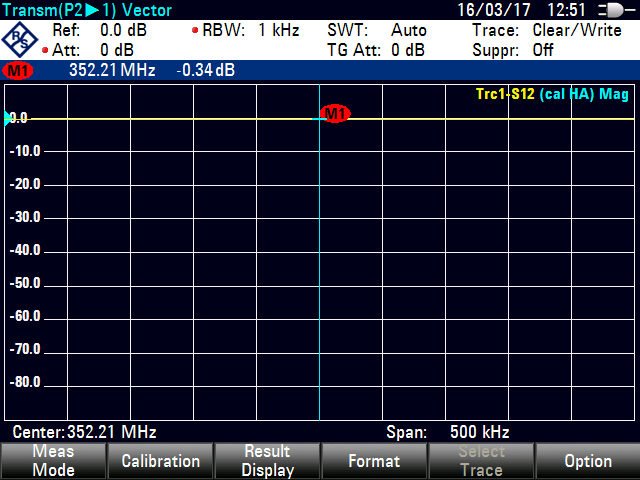} \protect\caption{The insertion loss measurement of the whole RF transmission line.  \label{fig:wholeS21}}
\end{figure}

\begin{figure}[htbp]
\centering{}\includegraphics[width=0.8\columnwidth]{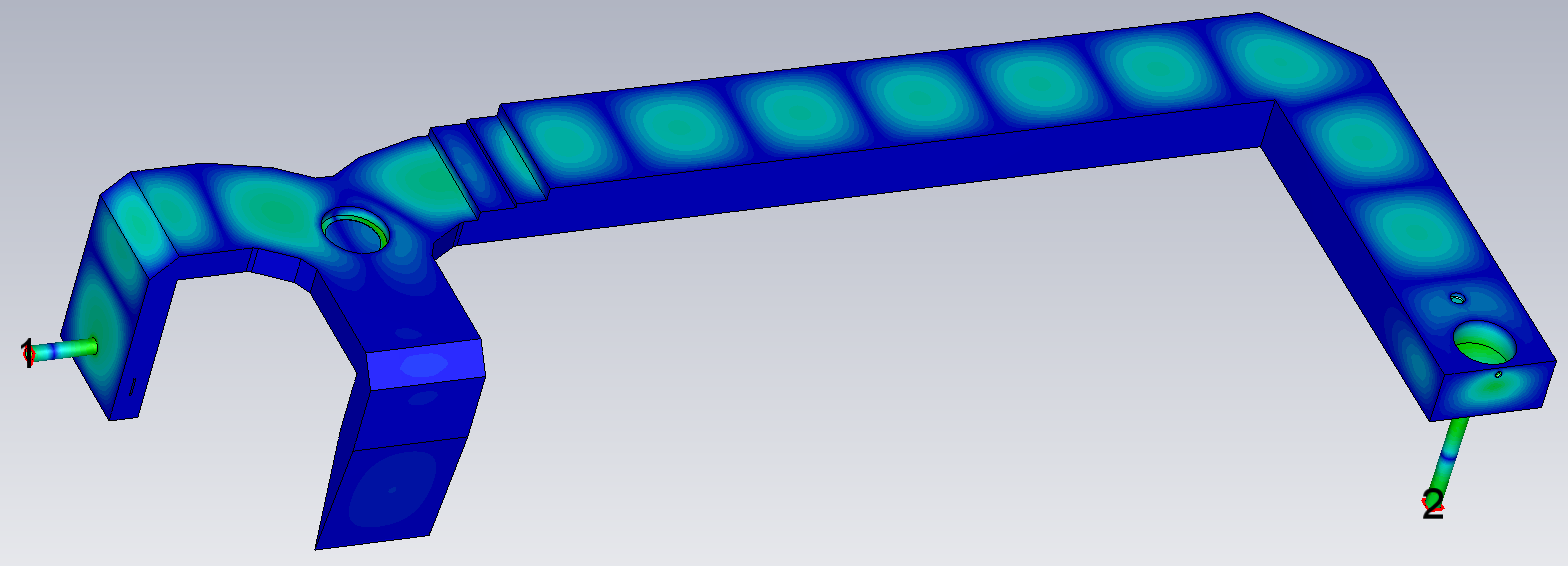} \protect\caption{The electric field pattern of the full RF transmission line. \label{fig:cstline}}
\end{figure}

After all the components were assembled, the full transmission line (with circulator) was fed
with low power RF signals to measure the S-parameters. The insertion loss for the full line is measured as -0.34 dB which is corresponds to 7.5\% power loss (Fig.\ref{fig:wholeS21}). Simulated electric field distribution of the whole RF transmission line is shown in Fig.\ref{fig:cstline}.

\section{High Power Measurements}

\begin{figure}[htbp]
\centering{}\includegraphics[width=0.7\columnwidth]{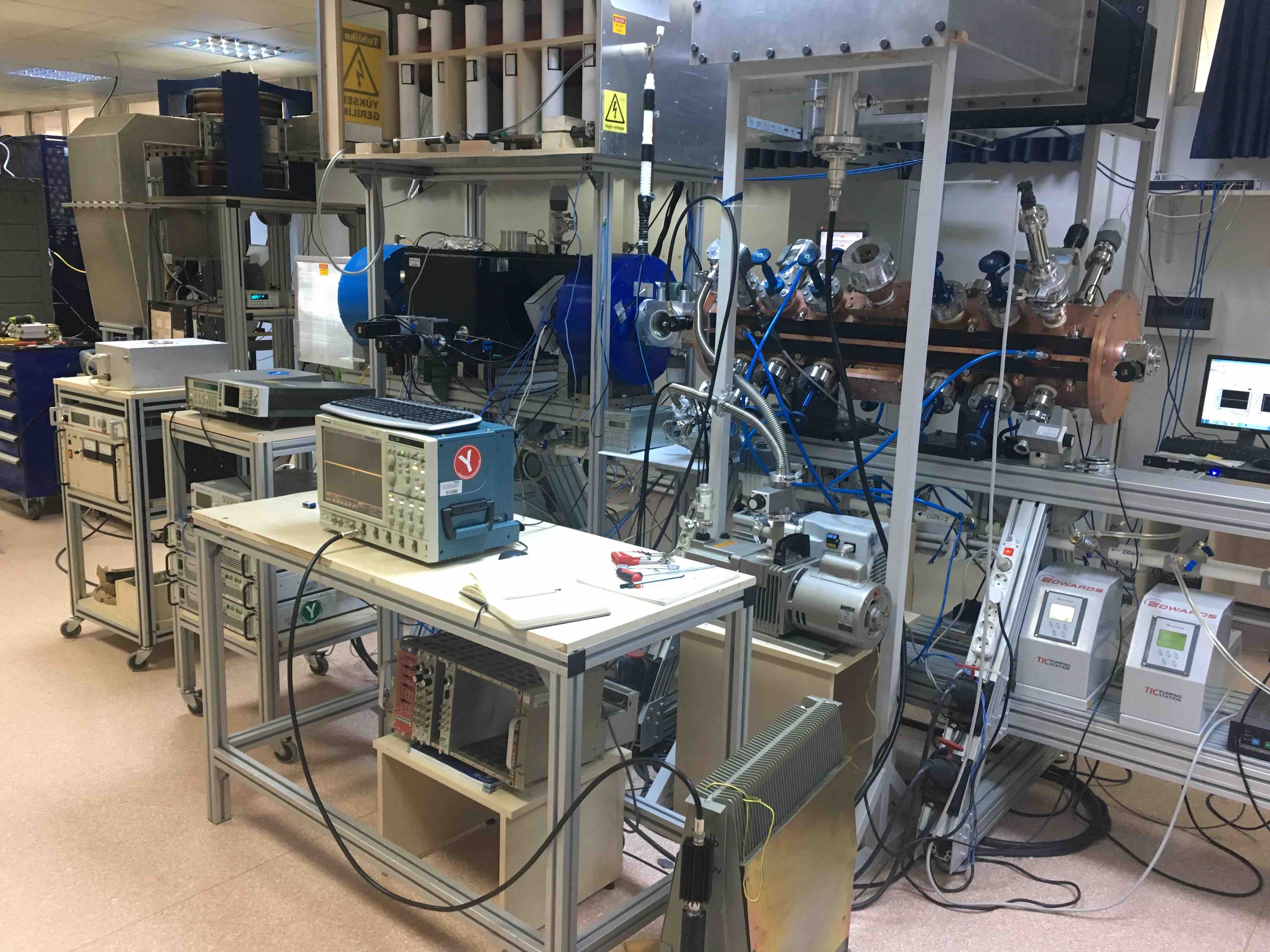} \protect\caption{High power test setup at the end of the RF transmission line.  \label{fig:HPtests}}
\end{figure}

The high power tests of the RF system were performed within three phases. First, at the SSA output, second at the tetrode output, and third at the end of the RF transmission line. The performance of the RF PSU was measured with a -60 dB RF load and a oscilloscope (Fig.\ref{fig:HPtests}). All high power measurements were done with calibrated coaxial cables and attenuators. First, high power tests were performed at the SSA output with the previously mentioned setup. A 5.8 kW RF power was measured with a pulse length of 100 $\mu$s. Second, the same setup was used to calculate the tetrode's amplification gain through feeding the RF tube with SSA power. A 15.9 kW output RF power was measured with an input power of 780.3 W. Therefore the gain of the tetrode tube was calculated as 13.1 dB. With this gain the tetrode is able to produce a 118.9 kW RF power with 5.8 kW RF input power. Unfortunately tetrode's output RF power at this power level could not measured due to the restriction of maximum power capability of the existing RF load. After the gain measurements, the tetrode output was connected to the RF transmission line. As the third phase, the high power tests were performed at the end of the RF transmission line. An SSA power of 2.4 kW was fed to the tetrode and the amplified power was measured at the end of the RF transmission line as 47.4 kW. Regarding a power loss of 7.5\% for the whole transmission line, the output power of the tetrode was calculated as 50.9 kW and the gain as 13.3 dB. Accordingly with this gain the RF PSU is enough to provide the required RF power to the RFQ cavity with a safety factor of 1.8 in case of reduction in the quality factor (Q) of the cavity.

\section{RF Interlock System}
\begin{figure}[htbp]
\centering{}\includegraphics[width=0.9\columnwidth]{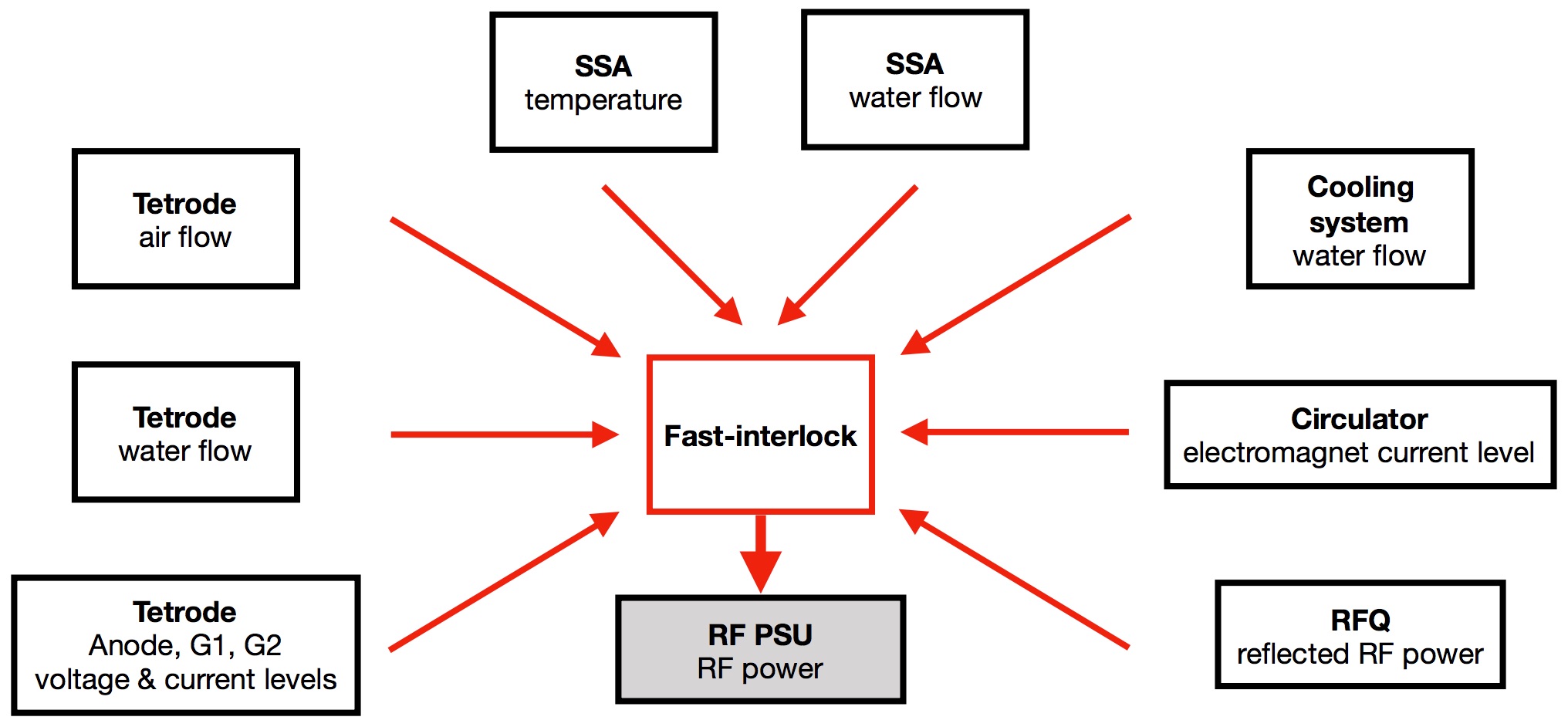} \protect\caption{Fast-interlock diagram of the RF system.  \label{fig:interlock}}
\end{figure}

The RF system is protected by predefined interlock conditions (Fig.\ref{fig:interlock}).
On the power supply side, the temperature of the SSA
is continuously monitored, as well as the availability of the water cooling.
For the tetrode amplifier, both water and air cooling requirements are imposed
for activating the second stage. Tube's anode, G1 and G2 voltages and currents are monitored and in case of unwanted level change the fast-interlocks (100 ns) protect the system. Additionally, the PSU finite state machine requires
the "ON" signal from the independent measurement of the water flow rate (should be
higher than 15 l/m) and from the circulator biasing electromagnets. In case of
any failure, the PSU initiates an automated shutdown procedure. The cooling water pressure and temperature are also continuously displayed and saved to a file. The reflected RF power is continuously monitored with a directional coupler just before the RFQ to provide a fast shutdown signal to the RF PSU in case of a value above a user defined threshold.

\section{Conclusions and Prospects}
The RF power supply and the RF transmission line for the SPP proton
beamline are designed, built and installed as a combined effort between
the SPP team and local manufacturers. The measurement results (power and S-parameters, etc.)
on the RF transmission line correspond well to the design expectations.

The educational and training outcomes are immense since the accelerator physicists and engineers were able to design, to help the manufacturers to produce their design, and finally
to compare the finished product to the concept they had initially. Moreover,
the manufacturers understand the expectations concerning the level of precision and details
for accelerator component production. And finally the fact that, in Turkey, there
is now a set of physicists, engineers and manufacturers who can speak
the same language, work on a complex product from its design to its
delivery could be a big regional asset. 

\section*{Acknowledgments}

\indent
The authors are grateful to A. Tanrikut for useful comments. This project is supported by the TAEK, under project No.~A4.H4.P1.


\end{document}